\documentclass[twocolumn,aps,superscriptaddress,pre]{revtex4}

\usepackage{amsmath,amssymb,graphicx}
\usepackage{algorithmic}
\usepackage{enumerate}
\usepackage{times}
\usepackage{color}
\usepackage{soul}
\definecolor{yblue}{rgb}{0.06, 0.3, 0.57}
\usepackage[pdftex]{hyperref}
\hypersetup{colorlinks=true,linkcolor=blue,citecolor=blue,urlcolor=blue}

\begin{document}

\title{%Dark soliton 
Ring dark solitons in three-dimensional Bose-Einstein condensates}

\author{Wenlong Wang}
\email{wenlongcmp@gmail.com}
\affiliation{Department of Physics, Royal Institute of Technology, Stockholm, SE-106 91, Sweden}

\author{P. G. Kevrekidis}
\email{kevrekid@math.umass.edu}
\affiliation{Department of Mathematics and Statistics, University of Massachusetts,
Amherst, Massachusetts 01003-4515 USA}

\author{Egor Babaev}
\affiliation{Department of Physics, Royal Institute of Technology, Stockholm, SE-106 91, Sweden}

\begin{abstract}
  In this work we present a systematic study of the three-dimensional
  extension of the
  ring dark soliton examining its existence, stability, and dynamics in isotropic harmonically trapped Bose-Einstein condensates. Detuning the chemical potential
  from the linear limit, the
  ring dark soliton becomes unstable immediately, but can be
  fully stabilized by an external cylindrical potential. The ring has a
  large number of unstable modes which are analyzed through
  spectral stability analysis. Furthermore, a
  few typical destabilization
  dynamical scenarios are revealed with a number of interesting vortical
  structures emerging  such as the two or four coaxial parallel vortex rings.
  In the process of considering the stability of the structure, we also
  develop a modified version of the degenerate perturbation theory method
  for characterizing the spectra of the coherent
  structure. This semi-analytical method
  can be reliably applied to any soliton with a linear limit to explore
  its spectral properties near this limit. The good agreement of
  the resulting spectrum is illustrated via a comparison with the
  full numerical Bogolyubov-de Gennes spectrum. The application of the method to 
the two-component ring dark-bright soliton is also discussed.
\end{abstract}

%\pacs{75.50.Lk, 75.40.Mg, 05.50.+q, 64.60.-i}
\maketitle

\section{Introduction}
Bose-Einstein condensates (BECs) have been of intense interest
for more than two decades since their experimental realization
because they provide a controllable
playground for investigating macroscopic quantum
phenomena~\cite{becbook1,becbook2}. A major research track
within this broad theme concerns
the nonlinear properties of the solitary waves supported by the weakly interacting gas, sharing many similarities with nonlinear optics \cite{DSoptics}. In this context, a large variety of solitary waves has been studied in the context of BECs, ranging from dark solitons~\cite{Frantzeskakis_2010},
vortices and vortex lines~\cite{Alexander2001,fetter2}, vortex rings~\cite{komineas} to more complex vortical structures such as long-lived
knots \cite{PhysRevE.85.036306}.
% ,PhysRevA.99.063604}.
On the other hand, the creation of fermionic and multicomponent condensates is believed to be useful for understanding the high-temperature superconductivity \cite{Jin:FBEC}. On the practical side,
some of these structures such as the dark solitons have recently been proposed for use as qubits with remarkably long lifetimes \cite{DSqubits}.
Some additional three-dimensional (3D) structures in attractive and repulsive nonlinearities can be found (respectively), e.g., in Refs.~\cite{boris1,boris2}.

\begin{figure}%{r}{0.5\textwidth}
\includegraphics[width=\columnwidth]{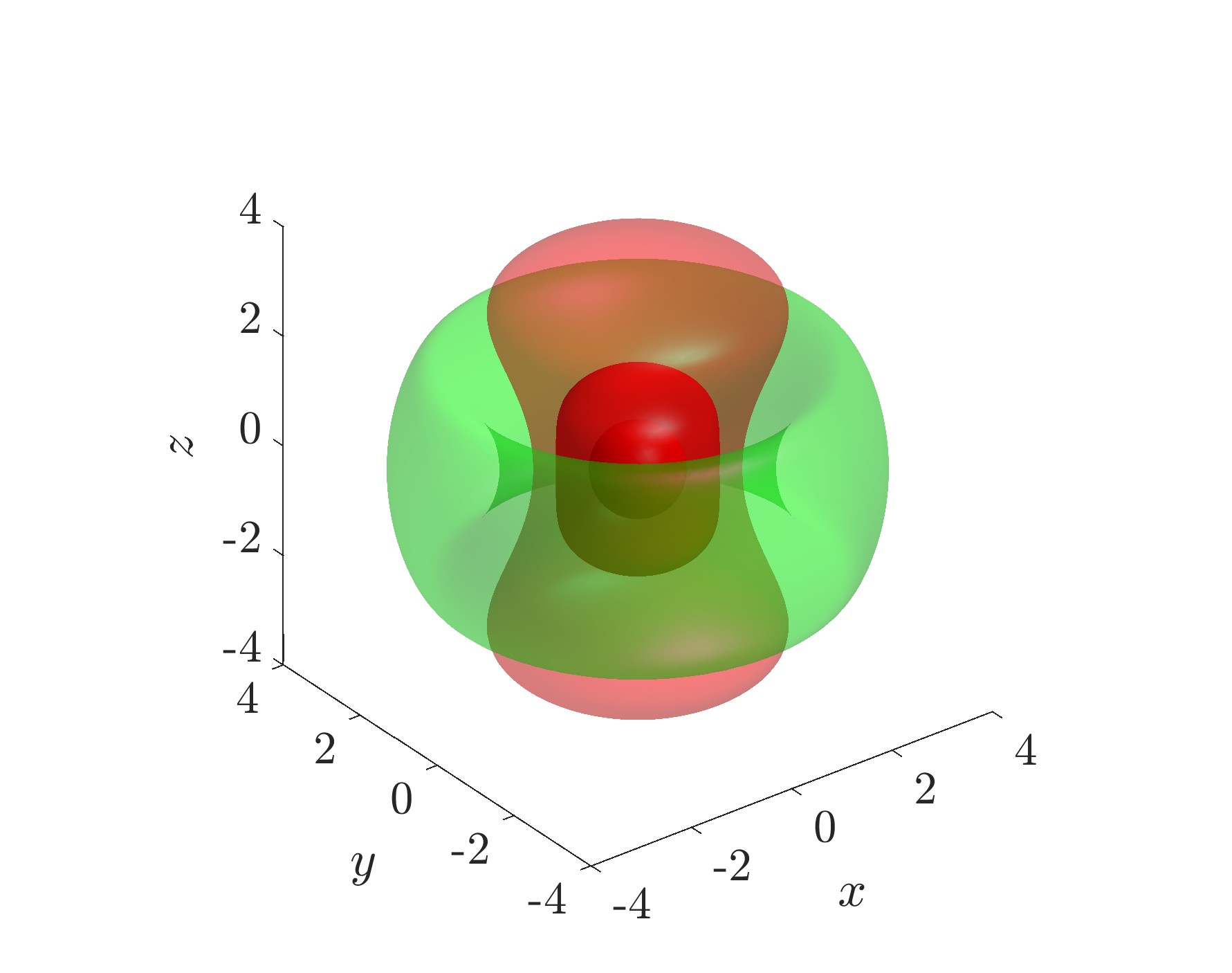}
\caption{Density contour and phase profile of a stationary ring dark soliton in a spherical harmonic trap of frequency $\omega=1$ at chemical potential $\mu=16$ (Eq.~(\ref{GPE})). There is a ring of vanishing density separating two segments with phase difference of $\pi$ shown as two different colours (red and green). Note the ring is not vertically straight, but bends. The state is generated from numerical computation.
}
\label{DSR}
\end{figure}

The main purpose of this work is to present a systematic study of
a prototypical soliton among the relevant coherent structures, namely
the ring dark soliton (RDS) but in the present setting as
a genuinely 3D structure. The RDS has been intensively studied in 2D in both one- and two-component condensates \cite{Clark:DSR,DSR:PP,Wang:DSR,SKA:DSR,Xiaofei:DSR}. In 2D, the soliton has a ring density node with a phase difference of $\pi$ across the ring. It is then straightforward to imagine the 3D RDS as a cylindrically extended
version of the 2D RDS along the axis of the ring. There is therefore a cylindrical-like surface of the (dark) node, again with a phase change of $\pi$ across the nodal surface. In practice,
for a fully 3D trap,
the ring is numerically found to deviate from being
straight along the z-axis; rather,
it bends in a harmonic trap, except at exactly the linear limit.
This is due to the incongruence between the cylindrical symmetry of
the 2D pattern and the closer to spherical (or more accurately: ellipsoidal)
nature of the 3D condensate. 
A cross section perpendicular to the axis is a 2D ring, and the radius is larger at the edges and smaller at the center. See Fig.~\ref{DSR} for a density and phase profile of the RDS.

Despite its extensive analysis in the 2D case, including the recent
corroboration of its modes of vibration and even
its multi-component extension~\cite{Panos:RDB}, 
the RDS has not been studied in detail in three dimensions.
Instead, other structures, such as
the planar dark soliton (PDS) \cite{Wang:DSVR,Wang:DBS} and the spherical dark soliton (SDS) \cite{Wang:DSS,Wang:DBS} have been considered
due to their  symmetry.  The latter have been dynamically
recognized  as generating spontaneously structures such
as vortex rings and multi-ring generalizations
thereof~\cite{Anderson:DS,Ionut:VR,Wang:DSVR}.
In that vein, such dark patterns (which are
somewhat more straightforward to generate
via phase imprinting as e.g. Ref.~\cite{DBS2} and more
recently in Ref.~\cite{PhysRevLett.119.150403})
  can be useful in an additional way.
In particular, they may be useful as seeds that, through the manifestation
of dynamical instabilities, may lead to
vortical structures that may be otherwise difficult to generate experimentally.
%It is also often hard to predict a priori what vortical filaments one could get from their decay products.
For example, the SDS can break into six vortex rings forming a cubic shape,
i.e., a vortex ring ``cage''; see Refs.~\cite{Wang:DSS,Wang:DBS} for details. Our study reveals that the 3D extension of the RDS is also capable of generating interesting vortical states such as the coaxial parallel two or four vortex rings (VRs), among others. The RDS actually has a  large number of unstable modes, providing an ideal setting for vortical line and ring
filament generation. Despite these instabilities, we find the RDS can be fully stabilized near the linear limit through the insertion of a suitable
external potential.
In the high density, large chemical potential Thomas-Fermi limit, the PDS,
SDS and 3D RDS are all subject to transverse instabilities that lead to a break
up into vortical structures. It is relevant to note in passing that the PDS
and SDS may be stable near the linear limit of a parabolic trap
without the need for the presence of an
additional external potential.
%Therefore, similarly to the PDS and SDS, the
%3D RDS is also unstable in the high density Thomas-Fermi limit, but also can be dynamically  stabilized near the low density linear limit.

Furthermore, our study also offers an interesting additional
twist.
In a number of the above works, the linear limit degenerate perturbation
method (LLDPM) has been used as a scheme enabling the identification
of the coherent structure's spectral modes in the vicinity of the limit.
%unexpectedly enabled us to find explicitly the reason of the occasional failures of the linear limit degenerate perturbation method (LLDPM) \cite{Wang:DSVR} and more importantly motivated us to find a reliable solution.
%The LLDPM is a very useful tool for analysing stabilities and computing asymptotic stability spectrum near the linear limit for a stationary soliton, benchmarking also the full numerical Bogolyubov-de Gennes spectrum.
This is a relevant method both for comparing with the full
Bogolyubov-de Gennes (BdG) numerical spectral stability analysis,
but also importantly for identifying the modes responsible for the
different emerging instabilities. 
%It works remarkably well for many solitons like the planar and spherical dark solitons. But it was found soon afterwards it does not work for every soliton most noticeably the vortex ring, a fundamentally interesting soliton, and it was not clear where the problem is. The method also fails for the RDS, therefore, the failure has nothing to do with whether the soliton is real or complex. We realize that, from
However, for waveforms such as 
the RDS, the method is not as successful, presumably due to the
deformation of the symmetry of the structure when one departs from
the linear limit (going from straight i.e., cylindrical at the limit
to bent in the nonlinear case).
This naturally suggests the design of a procedure to fix this problem by replacing the analytical linear waveform with a numerically exact state
which should be suitably normalized for the theory.
Then, performing a few numerical steps (that account for the nonlinear
symmetry of the RDS) and using controlled extrapolations, one retrieves
the good agreement with the full numerical spectrum of the RDS.
This is an effectively ``hybrid'' method combining the theoretical
expectation of the dependence of the different modes with a few
numerical data points used to subsequently perform
the relevant extrapolation.
Our findings suggest that weak nonlinearity can play a critical
role in conjunction with the theory of the solution. 

Our presentation is organized as follows. In Sec.~\ref{setup}, we introduce 
the model and the various analytical and numerical methods.
Next, we present our numerical results in Sec.~\ref{results}. 
Finally, our conclusions and a number of open
problems for future consideration are given in Sec.~\ref{conclusion}.

\section{Model and methods}
\label{setup}

We first present the mean-field Gross-Pitaevskii equation, and the relevant linear waveforms of the RDS. Then we discuss the numerical methods used for stationary states, spectral stability and dynamics. Finally, we summarize the details of the linear limit degenerate perturbation method and of its proposed
modification.

\subsection{The Gross-Pitaevskii equation}

In the framework of mean-field theory, and for sufficiently low temperatures, the dynamics of a 3D repulsive BEC, 
confined in a time-independent trap $V$, is described by the following
dimensionless Gross-Pitaevskii equation (GPE)~\cite{becbook1,becbook2}
\begin{equation}
i \psi_t= -\frac{1}{2} \nabla^2 \psi+V \psi +| \psi |^2 \psi-\mu \psi,
\label{GPE}
\end{equation}
where $\psi(x,y,z,t)$ is the macroscopic wavefunction of the BEC and 
$\mu$ is the chemical potential (the $t$ subscript denotes
a temporal partial derivative).
In this work, we consider a harmonic trap of the form: 
\begin{equation}
V=\frac{1}{2} \omega_{\rho}^2 \rho^2+\frac{1}{2} \omega_z^2 z^2,
\label{potential}
\end{equation}
where $\rho=\sqrt{x^2+y^2}$, $\omega_{\rho}$ and $\omega_z$ are the
trapping frequencies along the $x$-$y$ plane and %in 
the vertical direction $z$, respectively. 
Note that the potential has rotational symmetry with respect to the $z$-axis.
In our numerical simulations, we focus on the fully symmetric (isotropic,
spherically symmetric) case with
$\omega=\omega_{\rho}=\omega_z=1$ unless specified otherwise. Nevertheless,
we keep the notation more general
%the numerical methods and the perturbation methods are also
due to the intended broader applicability to the case of
a general aspect ratio $\omega_z/\omega_{\rho}$.
Here, the density $|\psi|^2$,  length, time, and energy are measured, respectively, in units of $1/(4 \pi a a_{\rho}^2)$, $a_{\rho}$, $\omega_{\rho}^{-1}$, and
$\hbar\omega_{\rho}$, where $a$ and $a_{\rho}=\sqrt{\hbar/(m\omega_{\rho})}$
are, respectively, the $s$-wave scattering length and the harmonic oscillator 
length in the $x$-$y$ plane.

Stationary states of the wavefunction
%(so called standing waves)
%are given by $\psi=e^{-i \mu t} \psi_0(x,y,z)$ and thus
$\psi_0$ satisfy the
elliptic partial differential equation (PDE):
\begin{eqnarray}
-\frac{1}{2} \nabla^2 \psi_0+V \psi_0 +| \psi_0 |^2 \psi_0 = \mu \psi_0.
\label{stationary}
\end{eqnarray}
The dark soliton ring has a linear limit associated with the
energy eigenstate of the corresponding simple harmonic oscillator with chemical potential $\mu_0=7\omega/2$.
We denote the eigenstate indices of the linear (i.e., $|\psi_0|^2 \rightarrow
0$) problem
in each of the different directions
as $n_x$, $n_y$ and $n_z$. Then,
in the corresponding $\{|n_xn_yn_z\rangle\}$ basis of the 3D quantum harmonic oscillator  eigenstates in Cartesian coordinates, the wavefunction for the RDS assumes the form:
\begin{eqnarray}
\label{linearwave}
|\phi_{\rm{RDS}}\rangle_{\rm{linear}}&=&
\frac{1}{\sqrt{2}}\left(|200\rangle+|020\rangle \right) \\
&\propto& \left(\rho^2-1\right) e^{-\omega r^2/2}.
\end{eqnarray}
In this work, we use $\psi$ for a nonlinear state of the GPE, and $\phi$ for a normalized simple harmonic oscillator wavefunction. The latter is often convenient for expressing the linear limit and will be useful when applying the linear limit degenerate perturbation method in the next section. Note that interestingly, the RDS has the same linear limit chemical potential (eigenvalue)
as the SDS which takes a slightly more complicated form in this basis $|\phi_{\rm{SDS}}\rangle_{\rm{linear}}=
\frac{1}{\sqrt{3}}\left(|200\rangle+|020\rangle +|002\rangle \right)$. This is a
degeneracy induced by the spherical trap.
%The SDS no longer has a linear limit once the spherical symmetry is lifted,
%The spherical dark soliton turns into an ellipsoid if the isotropic potential is deformed into an ellipsoidal shape
While the SDS is no longer a linear eigenstate in the absence of
spherical symmetry,
the RDS is a linear eigenmode for any aspect ratio $\omega_z/\omega_{\rho}$ as long as the trap has the rotational symmetry along the $z$-axis.

The stability of a stationary state $\psi_0$ can be analyzed
through the spectral BdG method. In particular, we linearize using:
\begin{equation}
\psi(\vec{r},t)=\psi_0(\vec{r})+ a(\vec{r})\exp(\lambda t)+b^*(\vec{r})\exp(\lambda^*t),
\label{perturbation}
\end{equation}
where $a, b$ are small, independent complex perturbations. Substituting Eq.~(\ref{perturbation}) into %the equation of motion 
Eq.~(\ref{GPE})
and retaining up to linear terms in the expansion, the linear stability eigenmodes $(a,b)^T$
%=(a_0,b_0)^T \exp(\lambda t)$
with eigenvalues $\lambda$ are determined by the eigenvalue problem of the following matrix operator:
\begin{eqnarray}
M&=&
\begin{pmatrix}
-i(\mathcal{L}+2|\psi_0|^2) & -i\psi_0^2 \\
i\psi_0^{*2} & i(\mathcal{L}+2|\psi_0|^2)
\end{pmatrix},
\label{matrix}
\end{eqnarray}
where
\begin{eqnarray}
\mathcal{L} &=& -\frac{1}{2} \nabla^2 +V -\mu.
\end{eqnarray}

The treatment has been so far in three dimensions. However, such direct computations in three dimensions are rather expensive, especially for the eigenvalues. The reason we discuss these in three dimensions will become clear when we introduce the linear limit degenerate perturbation method in the next section. For now, we proceed to discuss how to compute numerical stationary states and the spectrum less expensively using the symmetries of the setup. Given the rotational symmetry of the RDS,
we compute both the stationary state and the spectrum using a cross section in the $\rho$-$z$ plane.
The spectrum is computed using basis expansions through the partial wave method \cite{Wang:DSS,Wang:DBS}; see also Ref.~\cite{10.1093/amrx/abr007}. The idea is to take Fourier expansions for the perturbations $a$ and $b$ in Eq.~(\ref{perturbation}),
observing that different azimuthal modes decouple to leading order. This allows us to compute eigenvalues for each azimuthal Fourier $m$-mode separately (eigenvalues of modes $m$ and $-m$ are complex conjugates) and the full 3D spectrum is the union of all the individual 2D spectra. The technique can also be generalized e.g. to two-component systems, again for solitons with rotational symmetries up to topological charges~\cite{Wang:DBS}. In our work, we have collected the spectrum using $m=0, 1, 2, 3, 4$ and $5$ \cite{Wang:DSS}; see also Ref.~\cite{10.1093/amrx/abr007} for a discussion of the relevance of selecting only the lower
$m$ modes. Our stationary state is computed using a finite element method
for the discretization of space and utilizing an iterative
Newton's method towards convergence. The linear state is used as an initial guess near the linear regime, and the solution is parametrically continued to higher chemical potentials. The states can be used for constructing 3D states using 2D cubic spline interpolations, and the 3D states are then used for dynamics. Our dynamics are integrated using the regular fourth-order Runge-Kutta method in the fully 3D
case.

%Here are some practice \cite{Wang:AI,Wang:AI2}.
%Here are some practice \cite{Wang:DSR,Wang:DSS,Wang:bifurcation}.
%Here are some practice \cite{Wang:hopfions,Wang:DBS,Wang:SO2}.

\subsection{Nearly linear limit degenerate perturbation method}

The linear limit degenerate perturbation
method~\cite{PhysRevA.62.053606,Wang:DSVR} is a technique for studying the spectrum near the low density/nonlinearity regime of a coherent
structure, such as the RDS.
The state emerges from this limit and the method allows to explore the
modification of its existence and stability properties as we detune
away from the limit. 
%It also provides a benchmark against the numerical spectrum.

At the level of the existence problem, the method uses the following
expansion:
\begin{eqnarray}
\label{ppsi}
\psi_0 &=& \sqrt{\epsilon}\phi_0 + \epsilon^{3/2}\phi_1 +..., \\
\mu &=& \mu_0 + \epsilon\mu_1 +...,
\label{psimu}
\end{eqnarray}
where $\psi_0$ is the nonlinear eigenstate and $\phi_0$ is the \textit{normalized} corresponding linear state, here the RDS state of Eq.~(\ref{linearwave}). $\phi_1$ and $\mu_1$ are perturbation terms. Using self-consistency, we find that $\mu_1=\int |\phi_0|^4 d^3x$.

In the realm of spectral BdG analysis,
near the linear limit both $|\psi_0|^2$ and $\mu-\mu_0$ are small.
This allows for a perturbative treatment of the eigenvalue problem of Eq.~(\ref{matrix}). More specifically, we separate the matrix to
\begin{eqnarray}
  M&=&M_0+M_I,
  \label{meth1}
\end{eqnarray}
where
\begin{eqnarray}
  \label{meth2}
M_0&=&
\begin{pmatrix}
-i\mathcal{L}_0 &0 \\
0 & i\mathcal{L}_0
\end{pmatrix}, \\
M_I&=&
\begin{pmatrix}
-i\left(2|\psi_0|^2-(\mu-\mu_0)\right) & -i\psi_0^2 \\
i\psi_0^{*2} & i\left(2|\psi_0|^2-(\mu-\mu_0)\right)
\end{pmatrix}. \nonumber \\
\label{meth3}
\end{eqnarray}
Here,
\begin{eqnarray}
\mathcal{L}_0 &=& -\frac{1}{2} \nabla^2 +V -\mu_0.
\end{eqnarray}
Now we note  that $M_0$ is a shifted (by $\mu_0$)
variant of the harmonic oscillator problem and hence all the eigenvalues and eigenstates are known exactly. $M_I$ represents the small correction
thereof near the linear limit.
Thus, all terms in $M_I$ are linear to order $\epsilon$. The eigenvalues of $M_I$ (excluding the factor $\epsilon$) together with the
detuning correction $\mu_1$ then provide a detailed spectrum near the linear limit in the regular framework of the degenerate perturbation theory.
More concretely, each $\mu$ corresponds, via $\mu_1$, to a perturbation strength $\epsilon$ via Eq.~(\ref{psimu}). Once $\epsilon$ is determined, so
are the eigenvalue corrections proportional to $\epsilon$.

The complete basis of $M_0$ consists of both ``up'' states and ``down'' states, of the form
$
\begin{pmatrix}
|n_xn_yn_z \rangle \\
0
\end{pmatrix}
$
and
$
\begin{pmatrix}
0 \\
|n_xn_yn_z \rangle
\end{pmatrix},
$
with eigenvalues $E_{|n_xn_yn_z \rangle}-E_{\rm{RDS}}$ and $E_{\rm{RDS}}-E_{|n_xn_yn_z \rangle}$, respectively, up to a factor of $-i$ and $E_\Lambda$ is the eigenenergy of the 3D %simple 
quantum harmonic oscillator for eigenstate $\Lambda$. %These eigenvalues are the full spectrum at the linear limit. 
The relevant states for eigenvalues 
near the spectrum of Im($\lambda$)=0, 1 and 2 are listed in Table~\ref{para}.
%In the case of Im($\lambda$)=0, the same states are given as they are
%neither up, nor down but degenerate with the RDS eigenvalue $\mu_0$.
Note that the eigenvalues of $M_0$ are all imaginary given the diagonal
form of $M_0$ and the Hermitian nature of ${\cal L}$.
%, or a linear state is dynamically stable.

\begin{table}
\caption{
Relevant states for eigenvalues near the spectrum of Im($\lambda$) $=0, 1$ and $2$ for the linear and nearly linear limit degenerate perturbation methods. Here $|mnp \rangle$ stands for eigenstates with all possible distinct permutations with the quantum numbers $m,n$ and $p$.
\label{para}
}
\begin{tabular*}{\columnwidth}{@{\extracolsep{\fill}} l c r}
\hline
\hline
Im($\lambda$)  & ``up'' states  & ``down'' states \\
\hline
$0$  & $|002 \rangle$ $|011 \rangle$ & $|002 \rangle$ $|011 \rangle$ \\
\hline
$1$  & $|003 \rangle$ $|012 \rangle$ $|111 \rangle$& $|001 \rangle$ \\
\hline
$2$  &\qquad $|004 \rangle$ $|013 \rangle$ $|022 \rangle$ $|112 \rangle$ \qquad & $|000 \rangle$ \\
\hline
\hline
\end{tabular*}
\end{table}

It was realized after developing the theory that the method seems to work particularly well for structures such as the planar and spherical dark solitons, but may not be well suited for others such as the vortex ring or,
for that matter, the RDS. When
this is the case, the method appears to give a mixture of modes that are in agreement or disagreement with the numerical spectrum, %It was quite mysterious what makes the difference. The method also fails for the RDS.
without a clear hint of the origin of such disagreement.
However, we find the RDS turns out to be especially
useful in that it finally provides a hint of the problem, as well
as a simple (somewhat empirical) way to fix it. The issue at hand
appears to be that a solitary waveform near the linear limit may \textit{differ
  subtly} from the linear limit. In other words, the linear waveform may not faithfully capture the essential features of the nonlinear waveform and
in particular the symmetries (or lack thereof) in the latter.
In our case, the RDS state $\phi_0$ in Eq.~(\ref{ppsi}) near the linear limit differs noticeably from the state $\phi_{\rm{RDS}}$ in Eq.~(\ref{linearwave}): the
former wavefunction is vertically straight and features cylindrical
symmetry, but once the RDS is in the nonlinear regime, it bends and
no longer possesses this symmetry.

We provide a hybrid way to systematically fix the problem,
utilizing the modified structure of the nonlinear pattern
in the immediate vicinity of the linear limit.
The method, nearly linear limit degenerate perturbation
method (NLLDPM), as we will see in the next section, yields very good
agreement with the numerical spectrum for chemical potentials $\mu$ close
to the bifurcation point. The idea is quite straightforward: we use a numerically exact waveform in Eq.~(\ref{ppsi}) instead of an analytical waveform exactly at the linear limit for taking matrix elements of
$|\psi_0|^2$ when constructing $M_I$. We do this in the neighbourhood of the linear limit and final results are obtained using extrapolations.
I.e., the method is hybrid in that it uses Eqs.~(\ref{meth1}-\ref{meth3})
from the theory combining them with a numerical input for $\psi_0$.
More specifically,
the method follows these concrete steps:
\begin{enumerate}
\item First, compute a stationary state near the linear limit, such as at $\mu=\mu_0+0.05$.
\item Renormalize this state such that it has norm $1$. This becomes
  the effective $\phi_0$ and its corresponding $\mu_1$ is computed.
\item By expressing $\psi_0$ in terms of $\phi_0$ and substituting
  $\mu-\mu_0=\epsilon \mu_1 + \dots$, a power of $\epsilon$ can be
  factored out from $M_I$ which can now be written as $M_I=\epsilon N_I$.
  The eigenvalues $\nu_i$ of the matrix $N_I$ are then computed.
%  Compute using this state the spectrum $\{\lambda_i\}$ (of $M_I$ without the perturbation factor $\epsilon$) and $\mu_1$.
\item We repeat these steps for a series of chemical potentials, but avoid being too close to the linear limit for numerical accuracy. For example, a good set could be $\mu=\mu_0+0.05,\ \mu_0+0.04,\ ...,\ \mu_0+0.01$. Using these data one
  can extrapolate each of the eigenvalues $\nu_i$ to the linear limit and use the theory beyond the specified values. The full spectrum near the linear limit can then be constructed using the eigenvalues of $M_I$ $\{\epsilon\nu_i\}$.
\end{enumerate}

The extrapolation for $\mu_1$ as a function of $\mu$ is straightforward.
On the other hand, there are two groups of running eigenvalues as $\mu$ changes:
this requires some care. One group has eigenvalues that are almost constant independent of $\mu$. These are the nontrivial eigenvalues we are looking for. The other group has eigenvalues that are rapidly suppressed as $\mu$ approaches the linear limit. These eigenvalues are trivial ones and should be discarded.
We emphasize that these ones are fairly easy to identify. One does extrapolations to the nontrivial eigenvalues and $\mu_1$.
%Since these values are well bounded with good trends, the extrapolation is well controlled.
We will provide more details on all of these aspects in the next section. Finally, our method is generic and broadly applicable to localized
structures emerging from the linear limit. In addition to the RDS, we have also applied the method to the vortex ring state which has a linear limit when $\omega_z/\omega_{\rho}=2$ and the two-component RDB soliton, and our method again yields excellent results. 

\section{Numerical results}
\label{results}

We first present the basic properties of the RDS, the waveforms and the stability spectrum, and then the  comparison of the proposed modification
with the full numerical BdG spectrum. By following the corresponding unstable
eigenmodes, we examine three typical destabilization scenarios for the dynamics.
Finally, we present some case examples of (partial or full)
stabilization of the RDS, including filling in a bright component and
using an external repulsive potential.

\subsection{Basic properties of the RDS}
\begin{figure}%{r}{0.5\textwidth}
\includegraphics[width=\columnwidth]{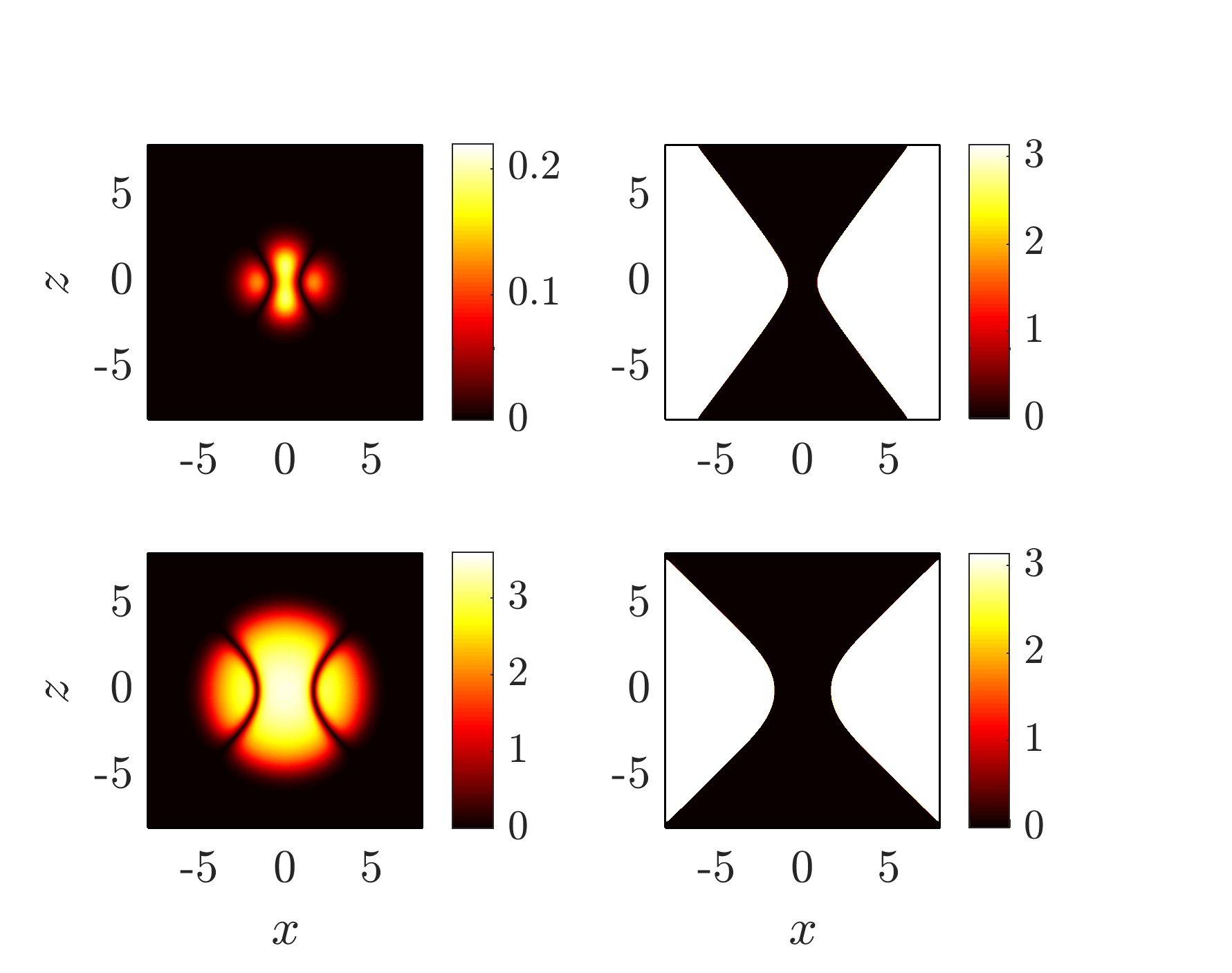}
%\put (-225,183) {(a) $\lambda=0.5$}
%\put (-145,183) {(b) $\lambda=1$}
%\put (-70,183) {(c) $\lambda=3$}
\caption{Two typical states of the RDS, one near the linear limit at $\mu=3.51$ (upper panels) and the other close to the Thomas-Fermi large-$\mu$
  limit at $\mu=12$ (lower panels).
  The left panels show $|\psi_0|$ and the right panels show the phase profile.
  The state has rotational symmetry with respect to the $z$-axis. Note that the RDS profile bends and departs from its cylindrical symmetry
  in a profound way immediately away from its linear limit. 
}
\label{states}
\end{figure}

We start by illustrating
typical RDS states in Fig.~\ref{states}. Since we have already showed the 3D profile
%and it has rotational symmetry,
we now show its cross sections in the $x$-$z$ plane. It is very clear
that the RDS has a phase jump of $\pi$ across the node and more importantly
that it bends. This starts in the vicinity of the linear limit and
progresses to a rather curved shape in the Thomas-Fermi (large $\mu$) limit. Note that the width of the RDS,
which also depends on the healing length, is narrower at the center and wider
at the edges as a result of the variation of the local density.

\begin{figure}%{r}{0.5\textwidth}
\includegraphics[width=\columnwidth]{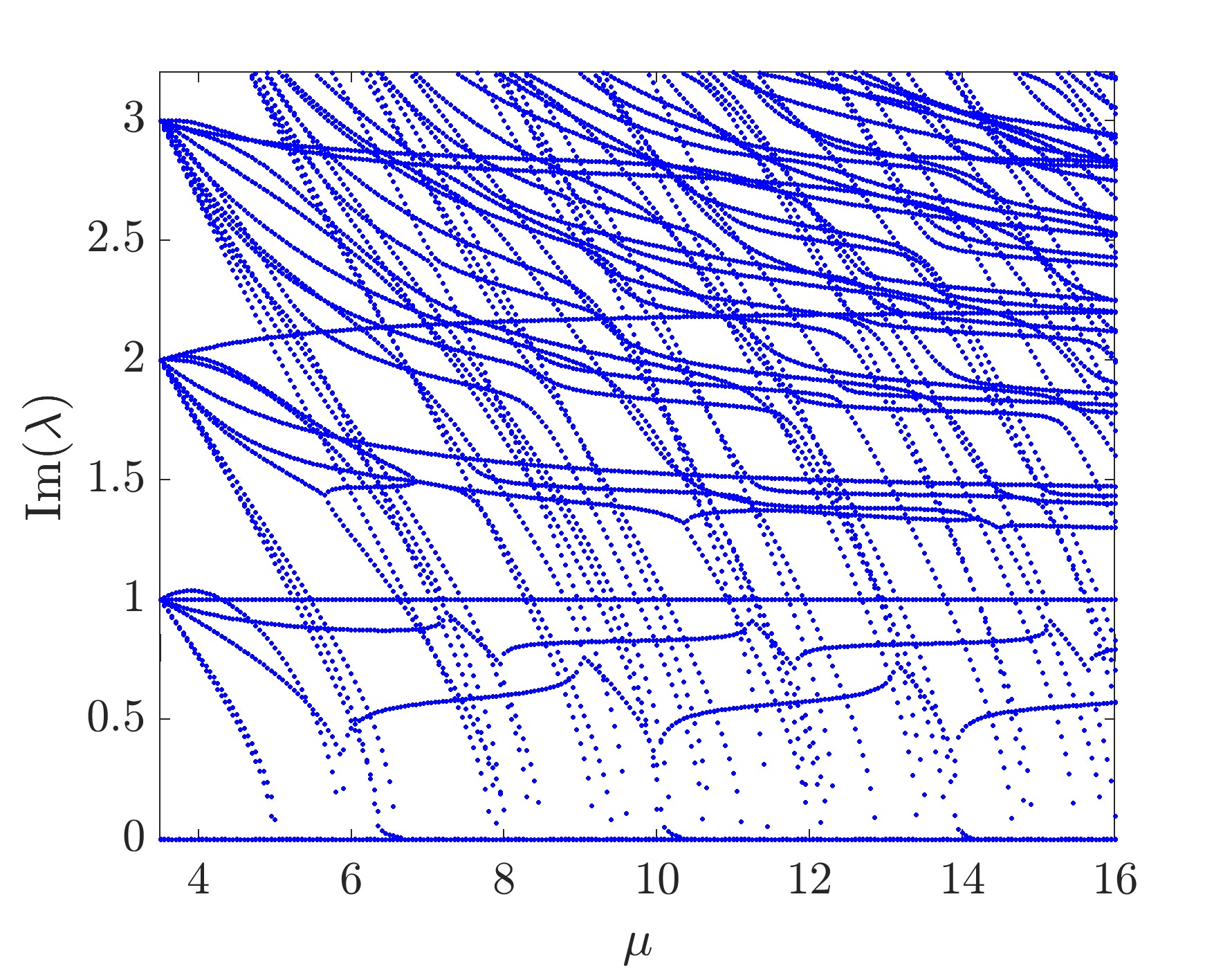}
\includegraphics[width=\columnwidth]{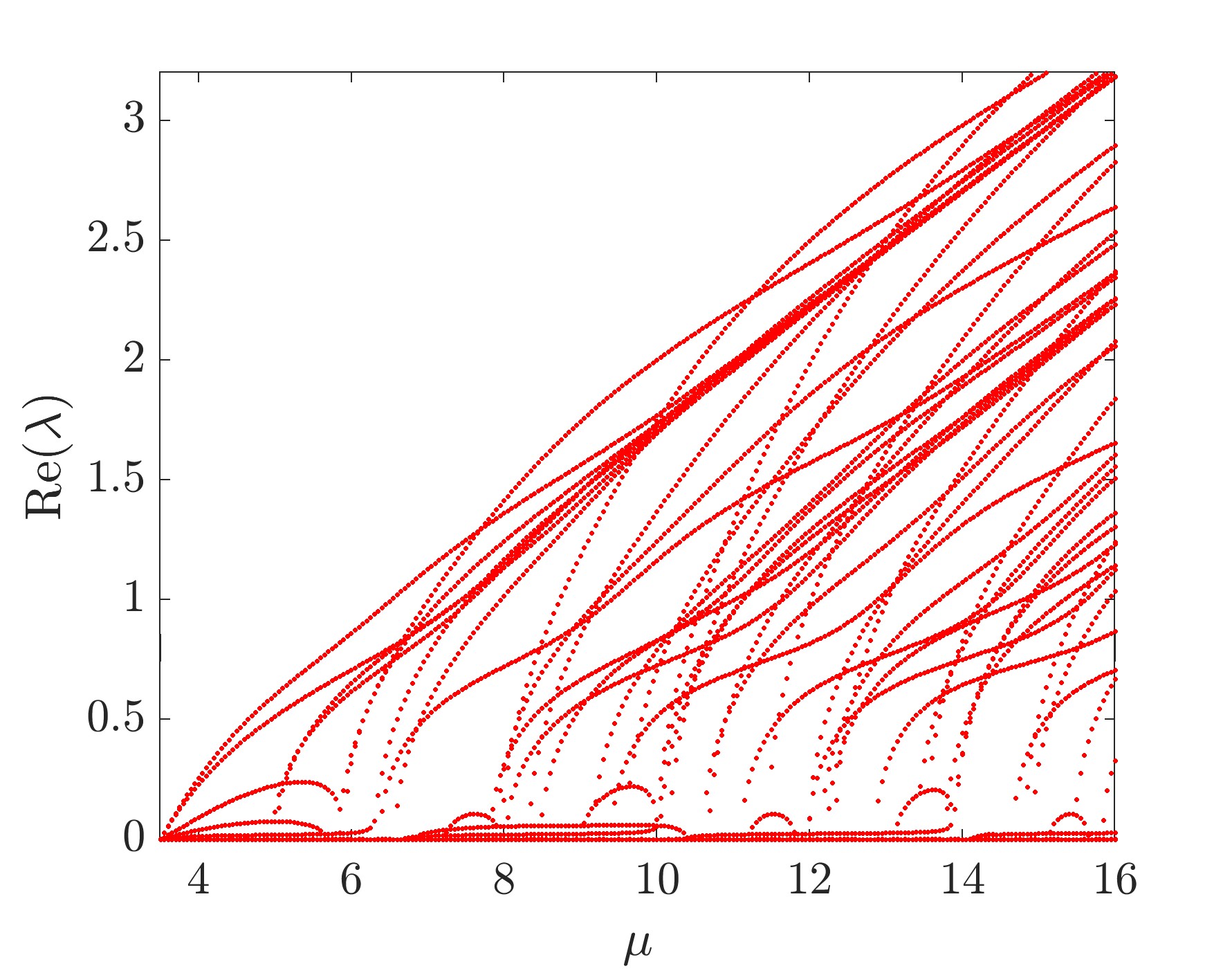}
\caption{The BdG spectrum of the RDS is particularly complex with many unstable modes. The blue and red points are for imaginary (stable) and real (unstable) eigenvalues, respectively. The RDS becomes unstable right from the linear limit and the maximum growth rate is an increasing function of the chemical potential.
}
\label{spectrum}
\end{figure}

\begin{figure}%{r}{0.5\textwidth}
\includegraphics[width=\columnwidth]{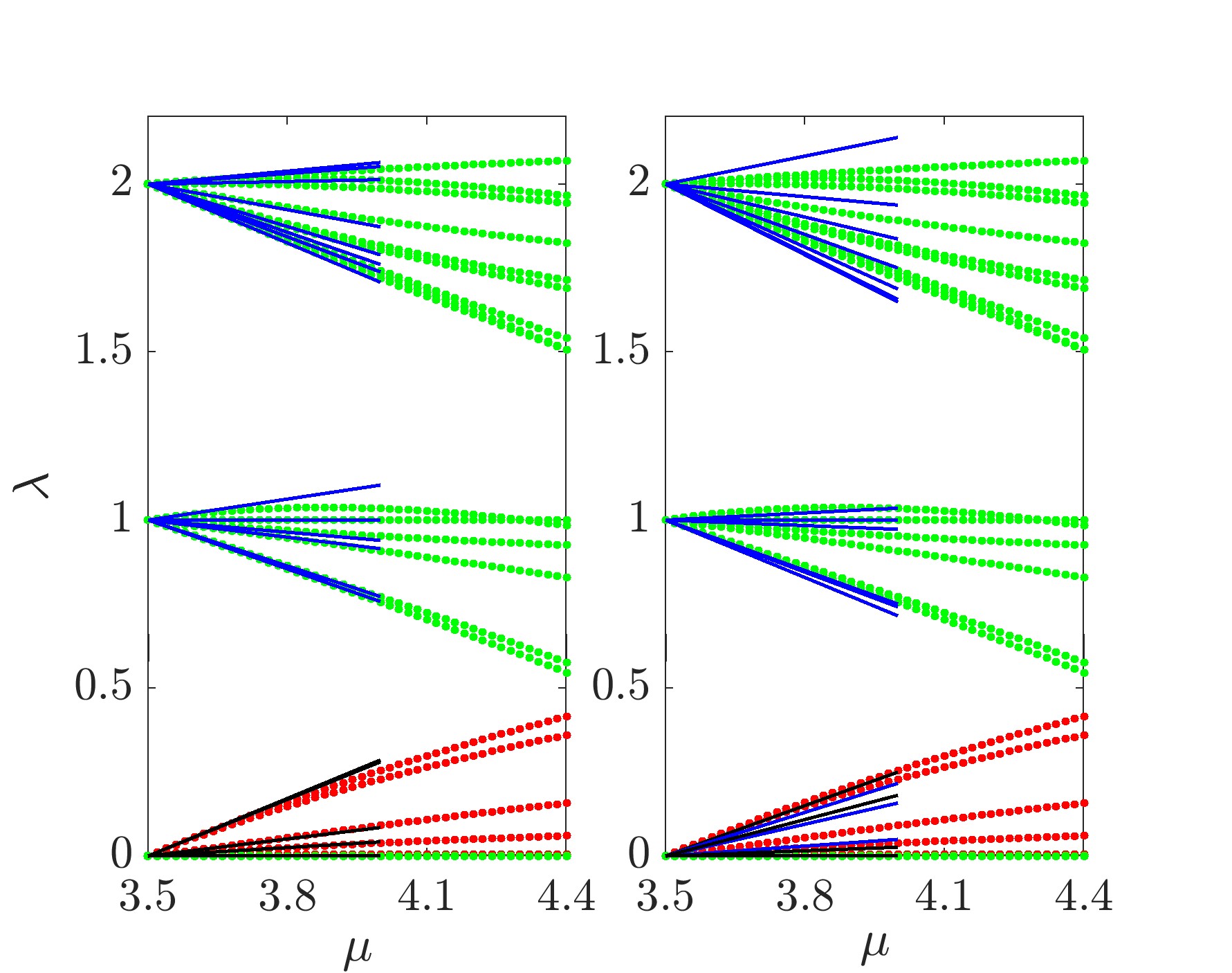}
\includegraphics[width=\columnwidth]{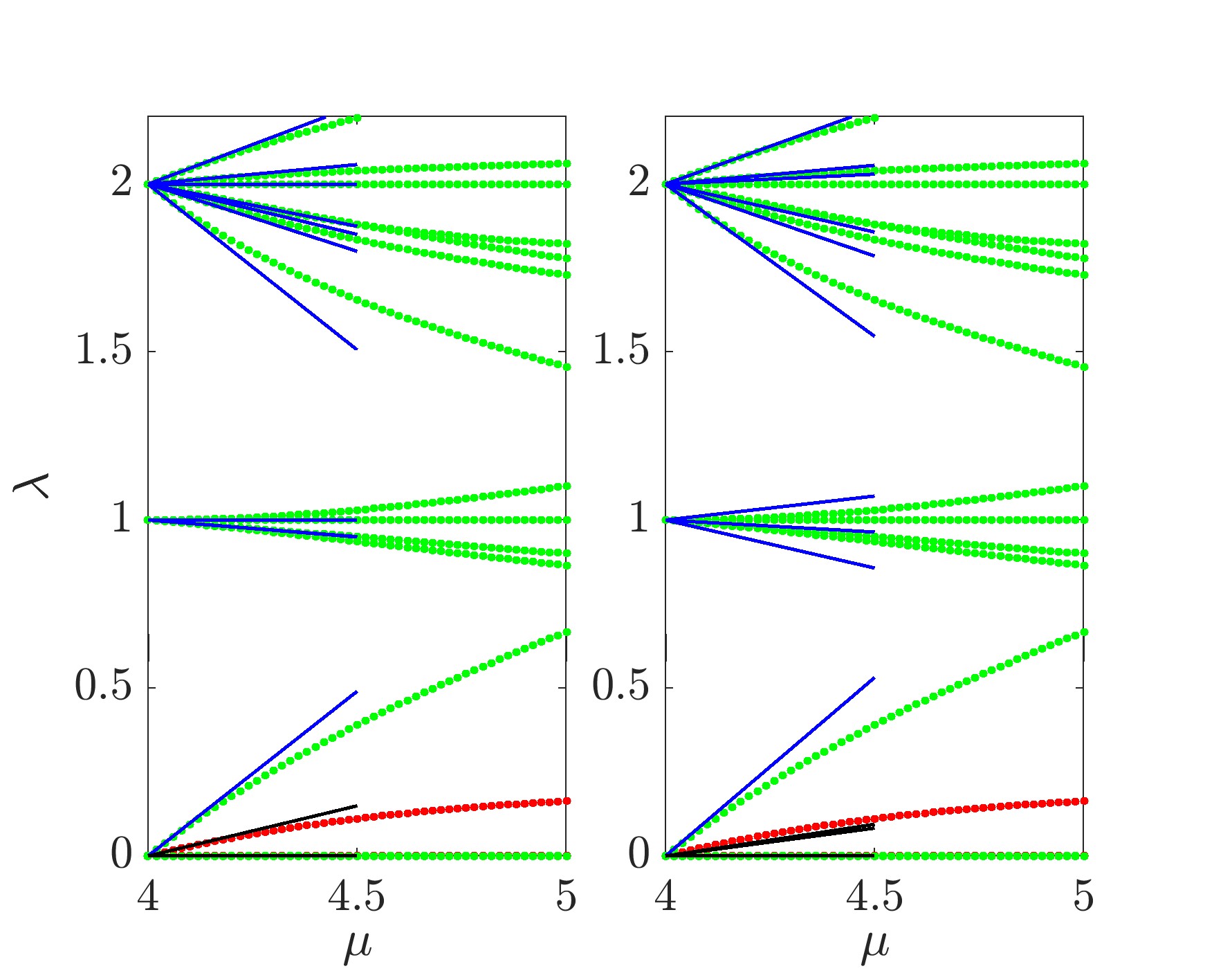}
\caption{Comparison of the NLLDPM (left) and LLDPM (right) with the full numerical spectrum (dots) near the linear limit, respectively. The top panels are for the RDS, while the bottom panels are for the VR. Red (black) are unstable modes and green (blue) are stable modes in the numerical computations
  (theoretical analysis). The LLDPM captures some modes correctly, but is generally not fully reliable both qualitatively and quantitatively.
  Careful inspection of the graph clearly indicates that
  the modified version NLLDPM matches excellently
  the numerical spectrum for all the relevant modes in the vicinity of
  the linear limit.
}
\label{DMP}
\end{figure}

The spectral stability analysis of the RDS is shown in Fig.~\ref{spectrum}. The RDS is unstable right from the linear limit. This is very similar to the 2D RDS \cite{Wang:DSR}, although the dominant  unstable mode is different as we discuss in the next section. The spectrum of the RDS is particularly complex
featuring numerous unstable modes. The instability growth
rate increases as a function of the chemical potential both in terms of the number of unstable modes and the maximum growth rate.
While the PDS and SDS maintain their symmetry as $\mu$ is increased,
the RDS becomes progressively more bent and deviates further from its
original cylindrical symmetry. This suggests that it is more prone
to resonances with different modes for different types
of unstable dynamics.

\begin{figure}%{r}{0.5\textwidth}
\includegraphics[width=\columnwidth]{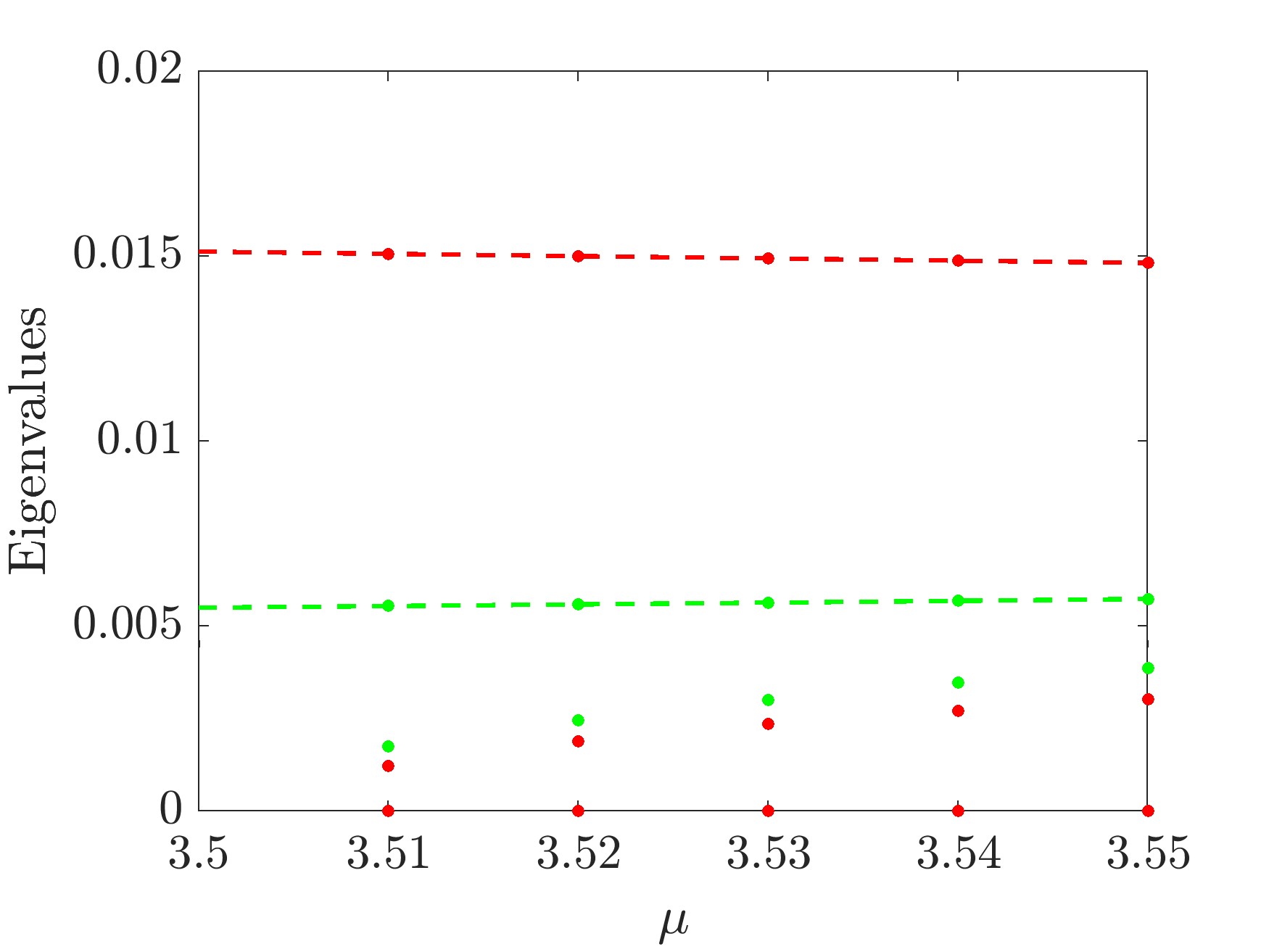}
\caption{Typical dependence of eigenvalues of the matrix
  $M_I$ without the factor of $\epsilon$ as a function of $\mu$ of the
  NLLDPM method for the RDS. We have selected two examples of trivial eigenvalues (one real and one imaginary) as shown at the bottom. These eigenvalues
  clearly bear the feature that they change rapidly and move toward zero. On the other hand, relevant eigenvalues do not change substantially. We show here also two examples of the largest real and imaginary eigenvalues along with the quadratic fits (shown by dashed lines).
%    Also do the fits have to be quadratic (in $\mu$ ?). Linear does not work ?
%  Also do note the addition re: dashed lines.}
}
\label{Eigenvalues}
\end{figure}

The comparison of our proposed (spectral) modification  and of
the full numerical spectrum near the linear limit is shown in Fig.~\ref{DMP}.
Careful inspection suggests that the LLDPM fails for both the RDS and the VR.
The failure of modes comes in two types: the theory may predict a
mode that does not exist or it may predict
a mode correctly but with a wrong slope.
As indicated, the modified version and the numerical spectrum are in
good agreement for all relevant eigenmodes. To get a better understanding of the
hybrid procedure, we present some examples for the extrapolation of $\mu_1$ (see Eq.~(\ref{psimu})) and some eigenvalues using the RDS as our example. Using five chemical potentials near the linear limit $3.55, 3.54, 3.53, 3.52$ and $3.51$, we obtain from the numerical stationary states $\mu_1= 0.02587, 0.02598, 0.02610, 0.02621, 0.02633$, respectively. We then extrapolate $\mu_1=0.02645$ using a quadratic polynomial fit. In contrast, the LLDPM gives $\mu_1=0.03175$ which differs by almost $20$ percent.
For the most unstable mode, the eigenvalues of the matrix $M_I$ without the factor of $\epsilon$ are $0.01482, 0.01488, 0.01494, 0.01500, 0.01506$ with an extrapolated value $0.01512$, and again the linear theory yields $0.01587$ which is off by about $5$ percent. This set of eigenvalues and the extrapolation as a function of $\mu$ along with other typical examples are shown in Fig.~\ref{Eigenvalues}. It is interesting to note the two sets of trivial modes at the bottom. Note that they run towards zero with substantial slopes; c.f. with the two other ones
(with nearly vanishing slope).
This allows us to identify irrelevant (spurious) modes in this procedure
and exclude them from the extrapolations. For the extrapolation method, we have used a quadratic polynomial fit. We have also tried to extrapolate using a cubic spline interpolation. The latter works for most cases, however, this method occasionally causes spurious bending phenomena. Therefore, we have selected a second order polynomial fit.

We emphasize that in as far as the case examples that we have
considered are concerned (including the RDS, VR, etc.),
the setup is robust. I.e., for a given coherent structure,
one can choose a properly selected set of chemical potentials and extrapolation techniques to apply the method. It thus appears that the assumption that the linear wavefunction can capture the essential features of the nonlinear waveform
has been a source of the limitations of the LLDPM. Properly
accounting for this via the NLLDPM provides significantly improved results.
%and when this is fixed using the nearly linear limit perturbation method, the problem is naturally solved.

It is well-known
%(already from early work, such as,
%e.g., in Ref.~\cite{Panos:DSR})
that the 2D RDS
nucleates an increasing number of vortex pairs as the chemical potential increases. As the spectrum of the 3D RDS differs signficantly from the 2D counterpart, we hence expect new modes of instabilities and different
decay products to arise. We now turn to the
corresponding dynamics in the next subsection.

\subsection{Instabilities of the RDS}
%In this section, we present RDS decay dynamics. As mentioned earlier, one of our major goals in
As explained previously, one of the principal motivations of our study
concerns the examination of  vortical patterns as a result of the destabilization of the RDS. As can be seen from Fig.~\ref{spectrum}, the dominant unstable mode changes with the chemical potential, and we have thus selected
a number of case examples for which we show the
decay dynamics at $\mu=5.5, 9$ and $13$.

\begin{figure}%{r}{0.5\textwidth}
\includegraphics[width=\columnwidth]{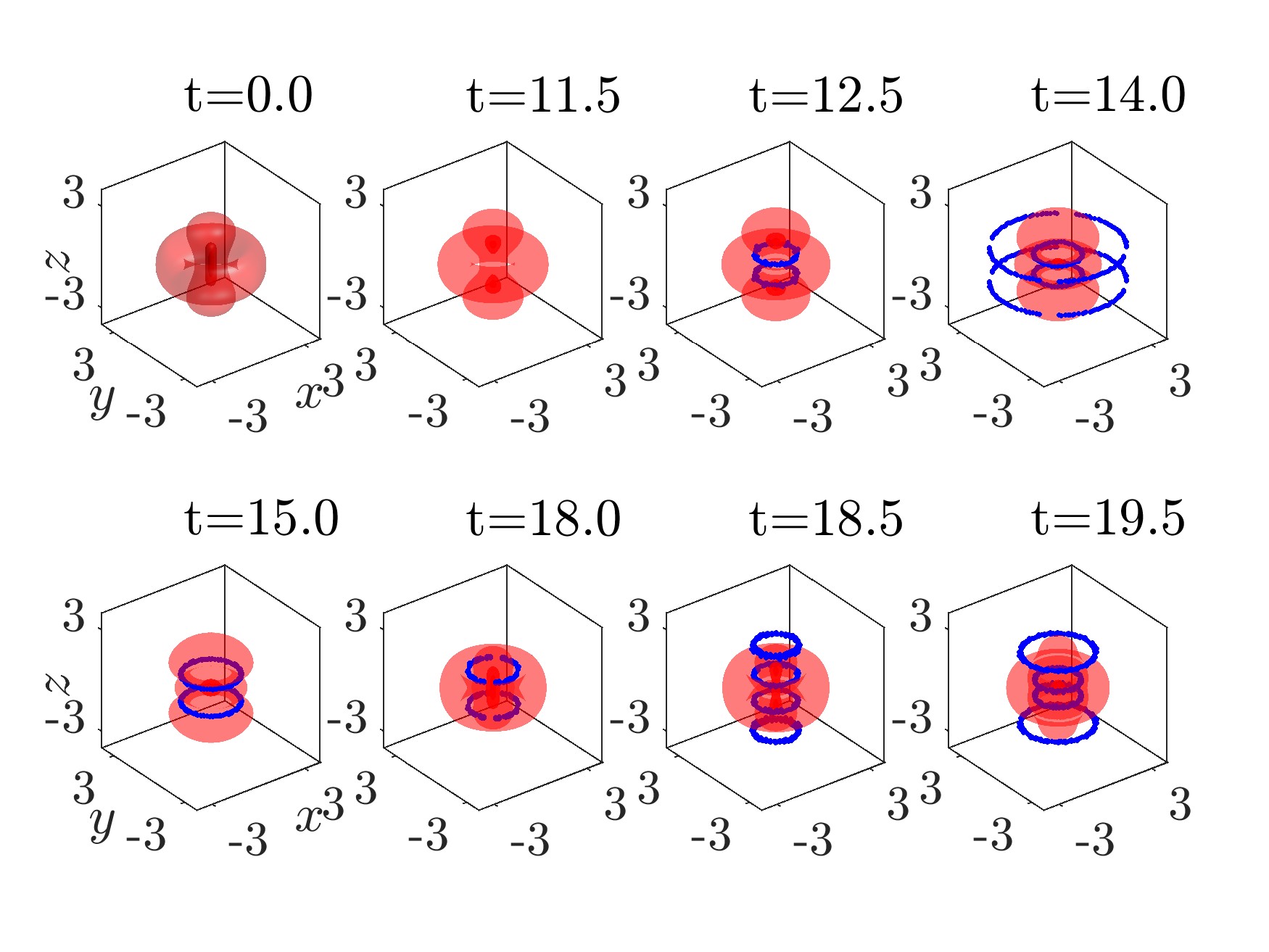}
\caption{Destabilization dynamics of the RDS at $\mu=5.5$. The RDS while unstable, can persist for quite a long time near the linear limit. The RDS firstly nucleates a vortex ring pair (VR2) and then another VR2 is created. The blue dots are the centers of the vortex rings. The inner pair rings move away from each other and the outer pair rings move toward each other and annihilate. The inner pair eventually reaches the edge of the condensate and turns around, and an additional pair of VR2 is induced in the process. See the text and the movie \cite{RDSmv1} for more details.
}
\label{RDSD1}
\end{figure}

The first relevant dynamics for $\mu=5.5$ is shown in Fig.~\ref{RDSD1}. The RDS persists until around $t=11.5$, after which it breaks up
nucleating two vortex rings (a state that we will call VR2 hereafter).
This is justified by the fact that the  RDS and the double planar dark
soliton (which we refer to as PDS2, stemming from the linear $|002\rangle$ state) have exactly the same linear limit chemical potential in the spherical trap. This mode therefore traces all the way back to the linear limit of the spectrum. The RDS hence is immediately unstable in the spherical trap, primarily as a result of the resonance with the PDS2. This produces the more stable state, namely the VR2 coming from the mixing of the RDS with PDS2 with a
relative phase of $\pi/2$; see also Ref.~\cite{Ionut:VR}.
It is relevant to highlight that in the 2D case,
the RDS firstly resonates with the double cross dark soliton (XDS2, stemming
from the 2D linear $|11\rangle$ state, again
with a $\pi/2$ relative phase between the 2D RDS and the XDS2). This
produces the relatively stable 
(aside from an oscillatory instability
for an interval of the chemical potential)
vortex quadrupole. The RDS should also be able to resonate with the 3D XDS2 (from the linear $|110\rangle$ state).
It is worthwhile to note that there are more than one unstable modes near the linear limit, however, this resonance is not the dominant one in three dimensions.

In the process of the dynamics, the originally produced VRs move outward
in the condensate, while the condensate generates an additional pair of
VRs
%The condensate very soon makes another pair of VR2 of a much larger size, making
leading to a total of four parallel vortex rings (a state that will hereafter
be referred to as VR4). These rings are not in equilibrium and hence
continue moving within the trap. The inner pair moves away from each other, while the outer pair rings moves toward each other leading to their
annihilation. The ``inner'' VR2 continuously moves outwards, reaches the edge of the condensate and turns around, corresponding to an oscillatory mode of the VR2. In the process, the VR2 firstly expands (around $t=15.0$) and then shrinks (around $t=18.0$). Remarkably, the pair induces two more rings into the condensate when the VR2 turns around, forming again a VR4 (around $t=18.5$). These rings are highly dynamical,
thus they do not settle in equilibrium positions but rather
continue featuring complex subsequent dynamics. We shall not pursue
these further and move on to the next unstable mode. See the relevant movie for more details of the dynamics \cite{RDSmv1}.

\begin{figure}%{r}{0.5\textwidth}
\includegraphics[width=\columnwidth]{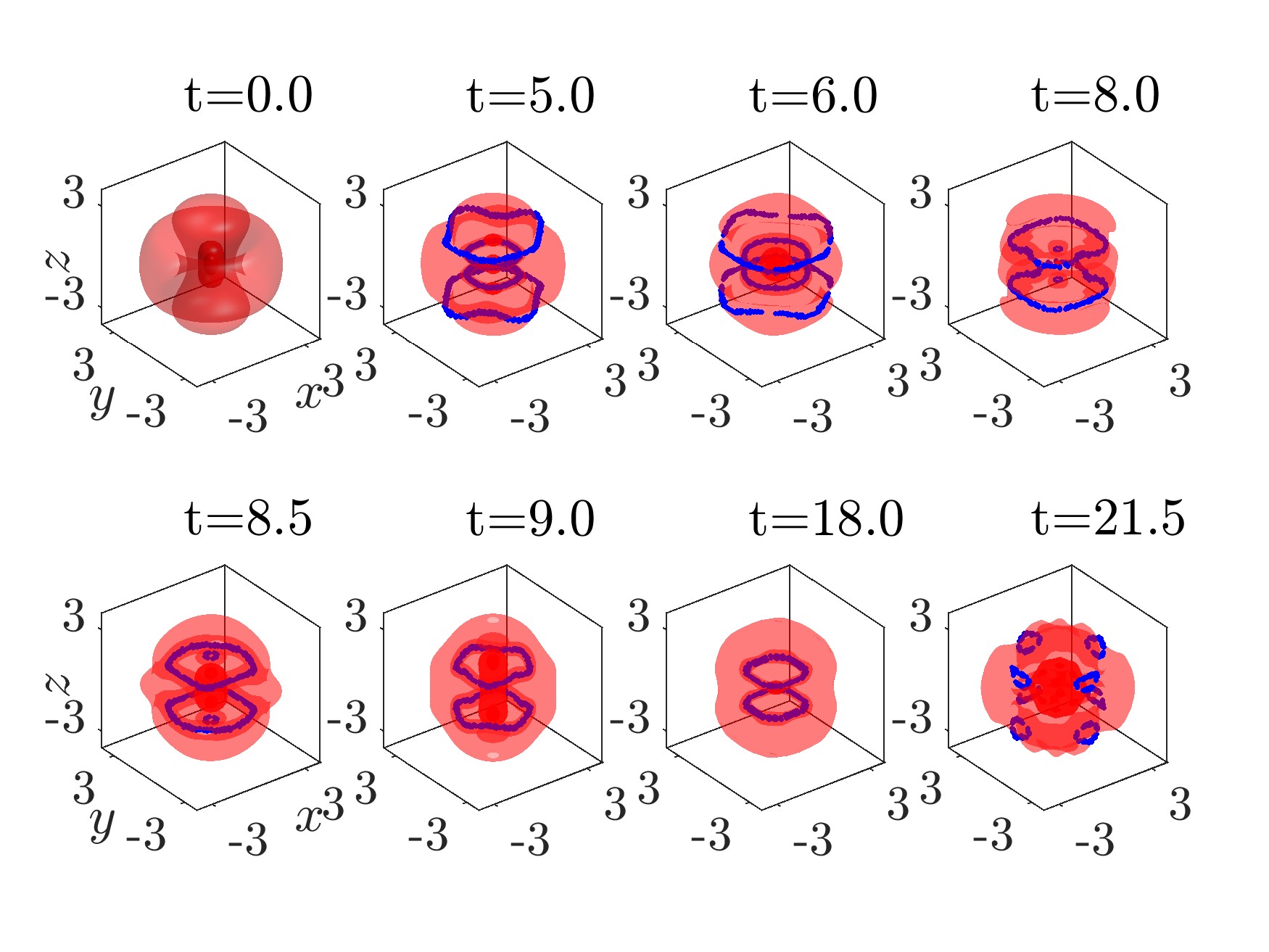}
\caption{Destabilization dynamics of the RDS at $\mu=9$. The RDS nucleates  an excited 4 vortex ring (VR4) state. The four rings move closer and collide (at about $t=7$) after which two rings are left. (larger rings of $t=8$). The two rings shrink, and then move outwards slightly, inducing two smaller rings (smaller rings of $t=8$). The small rings are highly mobile, and are quickly ejected out of the condenstate, leaving behind a long-lived VR2. The VR2 is, however, not fully stable and eventually
  breaks, forming a complex filament mixture such as a VR8 state at $t=21.5$. See the text and the movie \cite{RDSmv2} for more details.
}
\label{RDSD2}
\end{figure}

We have also considered the breakup dynamics for larger values of $\mu$,
associated with higher instability growth rates in Fig.~\ref{RDSD2}.
Here, the RDS breaks much faster, e.g., for  $\mu=9$, in agreement with the spectrum. The RDS nucleates an excited VR4 at around $t=5$. The VR4 is unstable, the inner smaller VR2 expands and the four rings move closer. After short and complex collisions, only one VR2 pair is left (the larger pair of $t=8$). The VR2 shrinks, and then moves outwards slightly, inducing another two new smaller rings (the smaller pair of $t=8$). The two small rings are, however, highly
mobile because of their small radius and are quickly ejected out of the condensate, leaving behind two excited but only weakly unstable rings at $t=9$. The VR2 is remarkably robust until about $t=18$. Finally, the rings move closer, undergo various breakings and reconnections forming a complex filament mixture. One interesting structure at $t=21.5$ is shown, where $8$ vortex rings (VR8) sit at the corners of a simple cube. It should be emphasized that this is also
a transient state which deforms quickly, i.e., it is not a stable stationary state. The observation of a long-lived VR2 is in agreement with our earlier work on the VR2 state \cite{Wang:VR}. The stationary VR2 is found to be almost stable at $\mu=9$ and is only fully stabilized at $\mu \gtrsim 15$. See the relevant movie for more details of the dynamics \cite{RDSmv2}.
%It is also much harder to stabilize the VR3 state, hence it is reasonable the VR4 state is not very stable at the chemical potential studied.

\begin{figure}%{r}{0.5\textwidth}
\includegraphics[width=\columnwidth]{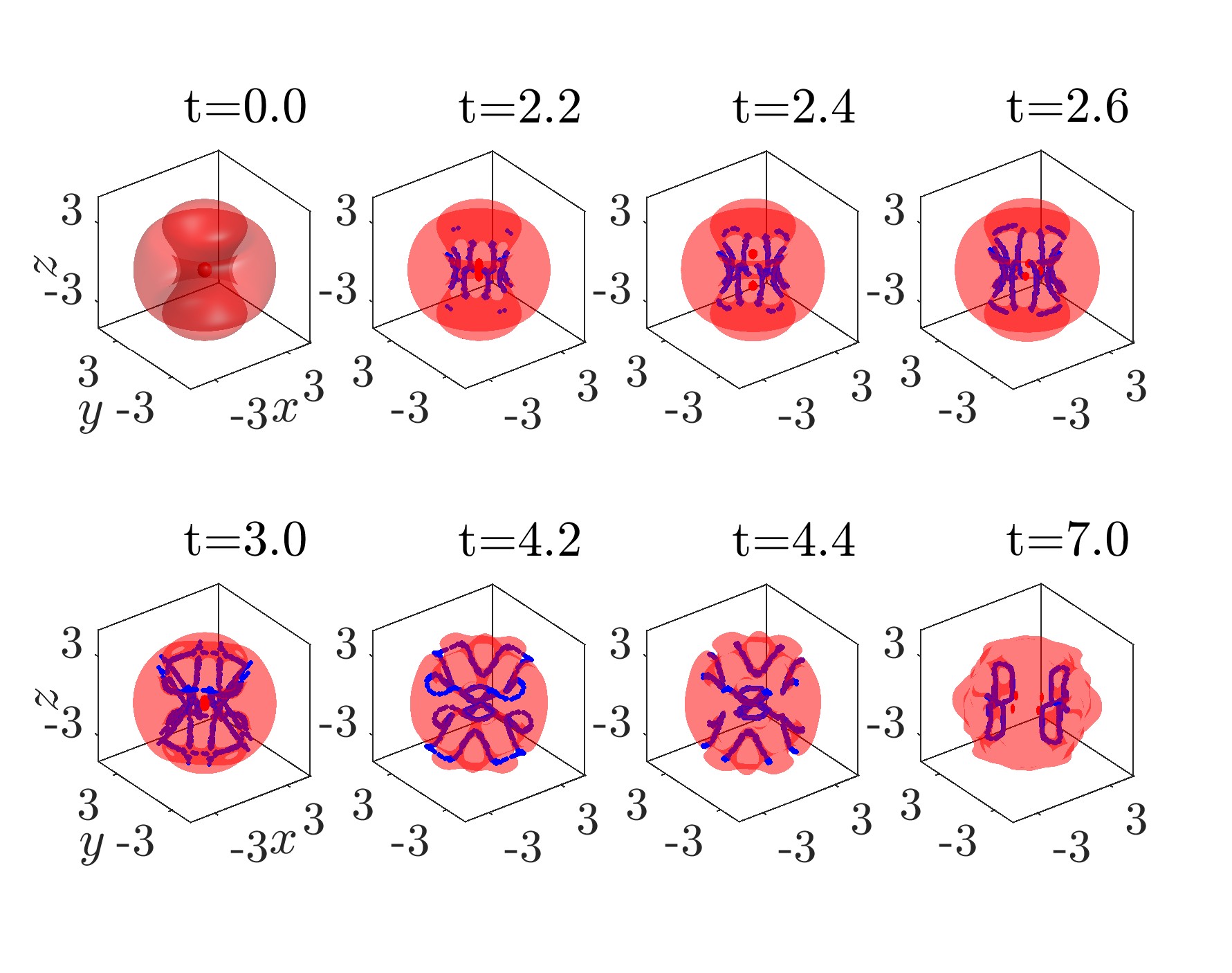}
\caption{Destabilization dynamics of the RDS at $\mu=13$. The RDS first nucleates to eight vortex lines (VL8). This is a state familiar from the perspective of the 2D RDS, and the $z=0$ cross section is a vortex octagon. The nucleation continues and forming a rather complex vortical structure at $t=3$, which bends inward and deforms to a highly excited VR4 state. The outer rings soon break to vortex lines and the inner rings run together and annihilate ($t=4.4$). The vortex lines have complex dynamics, forming e.g. the face to face VR4 state at $t=7$. See the text and the movie \cite{RDSmv3} for more details.
}
\label{RDSD3}
\end{figure}

Next, we discuss the destabilization dynamics at $\mu=13$ as
presented in Fig.~\ref{RDSD3}. The RDS breaks even faster, again in agreement with the spectrum. In this case, the core of the RDS firstly nucleates to eight, or four pairs of, parallel vortex lines (VL8). This is a familiar state from the perspective of the 2D RDS (where four vortex pairs may arise in the
destabilization dynamics). The $z=0$ cross section is hence a vortex octagon. The nucleation continues, creating four new pairs of vortex lines forming a cubic shape ($t=2.4, 2.6$). Then these vortex lines connect into a remarkably ordered but complex structure ($t=3$). The inner vortex lines bend inwards,
subsequently leading to a highly excited VR4 ($t=4.2$). Note the central
smaller pair of VRs is circular, while the outer pair is more excited. The outer pair then breaks into eight curved vortex line filaments, forming a cubic shape, and the inner pair moves closer. The inner pair soon annihilates at about $t=5.2$ (not shown). The outer vortex lines reconnect again into a highly excited VR2 (like the larger VR2 of $t=4.2$), and meanwhile forming and emitting four smaller rings out of the condensate. Then the VR2 reconnects to four VRs forming a square shape facing each other ($t=7$). The four rings move towards the center and suffer a complex collision. Subsequent dynamics (not shown here)
appear to still feature VR2 and VR4 states. See the relevant movie for
more details of the dynamics \cite{RDSmv3}. The spontaneous
  emergence
  of states such as the VR2, VR4 and VL8 is another one of the sources
of interest in the RDS structure. The latter structures are less
straightforward
to produce via other techniques such as phase imprinting, hence their
spontaneous emergence as a byproduct of the instabilities of the RDS
allows us to obtain insights on the interactions and dynamics of such (multi-)
vortex line and vortex ring states.

Finally, we present a two-mode analysis \cite{Wang:DSVR,Panos:TMA} of the bifurcations of the VR2, VR4 and VL8 states. This is a technique to compute approximately the bifurcation chemical potential of a new state arising
as a result of mixing from two relevant resonant solitary wave structures.
The relevant emergence is a result of a pitchfork bifurcation from
a ``parent'' branch (in our case the RDS one) which in the process
becomes destabilized (further, if it is already unstable).
Suppose the two linear limit stationary states are $\phi_1$ and $\phi_2$, with chemical potentials $\mu_1^0<\mu_2^0$. Then,
the theory~\cite{Panos:TMA} predicts the bifurcation chemical potential 
$\mu_c =\mu_1^0+(\mu_2^0-\mu_1^0)I_{11}/(I_{11}-I_{12})$, where the overlap integrals $I_{11}=\int |\phi_1|^4 d^3x$ and $I_{12}=\int |\phi_1\phi_2|^2 d^3x$. In our case, the RDS is always the state of the smaller chemical potential. The
(equal or) larger chemical potential states for the three cases are $|002\rangle, |004\rangle$ and a more complex cross dark soliton $\phi_{\rm{XDSm}} \propto \rho^m \cos(m\theta)\exp(-\omega r^2/2)$ with $m=4$ with a linear eigenenergy $(m+3/2)\omega$ \cite{Wang:DSS}, respectively. Evaluating the corresponding
integrals, we obtain the bifurcation chemical potentials
$\mu_c \approx 3.5, 6.25$ and $7.01$, in good agreement with the corresponding
critical points identified in
the spectrum $\mu_c \approx 3.5, 5.85$ and $7.85$.
The first of these bifurcations occurs already at the linear limit, giving
rise (in our spherical trap setting) to the VR2 state.
It is relevant to note, however, that the method is approximate (in that
it ignores higher order modes), and it becomes less accurate as $\mu_c$
deviates further from the linear
limit.
Nevertheless, it is reasonable to conclude that it is these
modes that resonate most strongly with the RDS, leading to the observed
bifurcations of new solutions out of the RDS coherent structure.
%making the dominant RDS unstable modes.

\subsection{Stabilization of the RDS}
Since the RDS is unstable for all values of the
chemical potential, we now look for ways to stabilize it. Our first attempt is to fill in a second component, forming a dark-bright soliton \cite{DBS1,DBS2}. Since this involves a two-component condensate, we briefly generalize the model we have used to two complex fields $\psi_1$ and $\psi_2$. Then, the dynamics with intra- and inter-component interactions is described as:
\begin{eqnarray}
i \frac{\partial \psi_1}{\partial t} &=& -\frac{1}{2} \nabla^2 \psi_1+V \psi_1 +(| \psi_1 |^2+| \psi_2 |^2) \psi_1  \nonumber \\
i \frac{\partial \psi_2}{\partial t} &=& -\frac{1}{2} \nabla^2 \psi_2+V \psi_2 +(| \psi_1 |^2+| \psi_2 |^2) \psi_2.
\end{eqnarray}
Here, we have in mind the setting of hyperfine states of, e.g., $^{87}$Rb
where the scattering lengths are nearly equal and hence it is a reasonable
approximation to set them equal as in the above mentioned, so-called
Manakov model~\cite{becbook1,becbook2}.
Here, standing wave solutions of the form
\begin{eqnarray}
\psi_1(\vec{r},t) &=& \psi^0_1(\vec{r})e^{-i\mu_1t} \nonumber \\
\psi_2(\vec{r},t) &=& \psi^0_2(\vec{r})e^{-i\mu_2t}
\end{eqnarray}
lead to the stationary equations:
\begin{eqnarray}
\label{SS1}
-\frac{1}{2} \nabla^2 \psi^0_1+V \psi^0_1 +(| \psi^0_1 |^2+| \psi^0_2 |^2) \psi^0_1 &=& \mu_1 \psi^0_1 \nonumber \\
-\frac{1}{2} \nabla^2 \psi^0_2+V \psi^0_2 +(| \psi^0_1 |^2+| \psi^0_2 |^2) \psi^0_2 &=& \mu_2 \psi^0_2,
\end{eqnarray}
where $\mu_1$ and $\mu_2$ are the chemical potentials of the first and second components, respectively. Here, it is straightforward to see there is a linear limit for the ring dark-bright (RDB) soliton at $\mu_1=3.5, \mu_2=1.5$, where the RDS is coupled to the ground state of the second component. We have similarly found the stationary states and computed the spectrum from the linear limit to a typical Thomas-Fermi limit $\mu_1=12, \mu_2=10$ with a linear ``trajectory'' in the ($\mu_1,\mu_2$) parameter space. 

\begin{figure}%{r}{0.5\textwidth}
\includegraphics[width=\columnwidth]{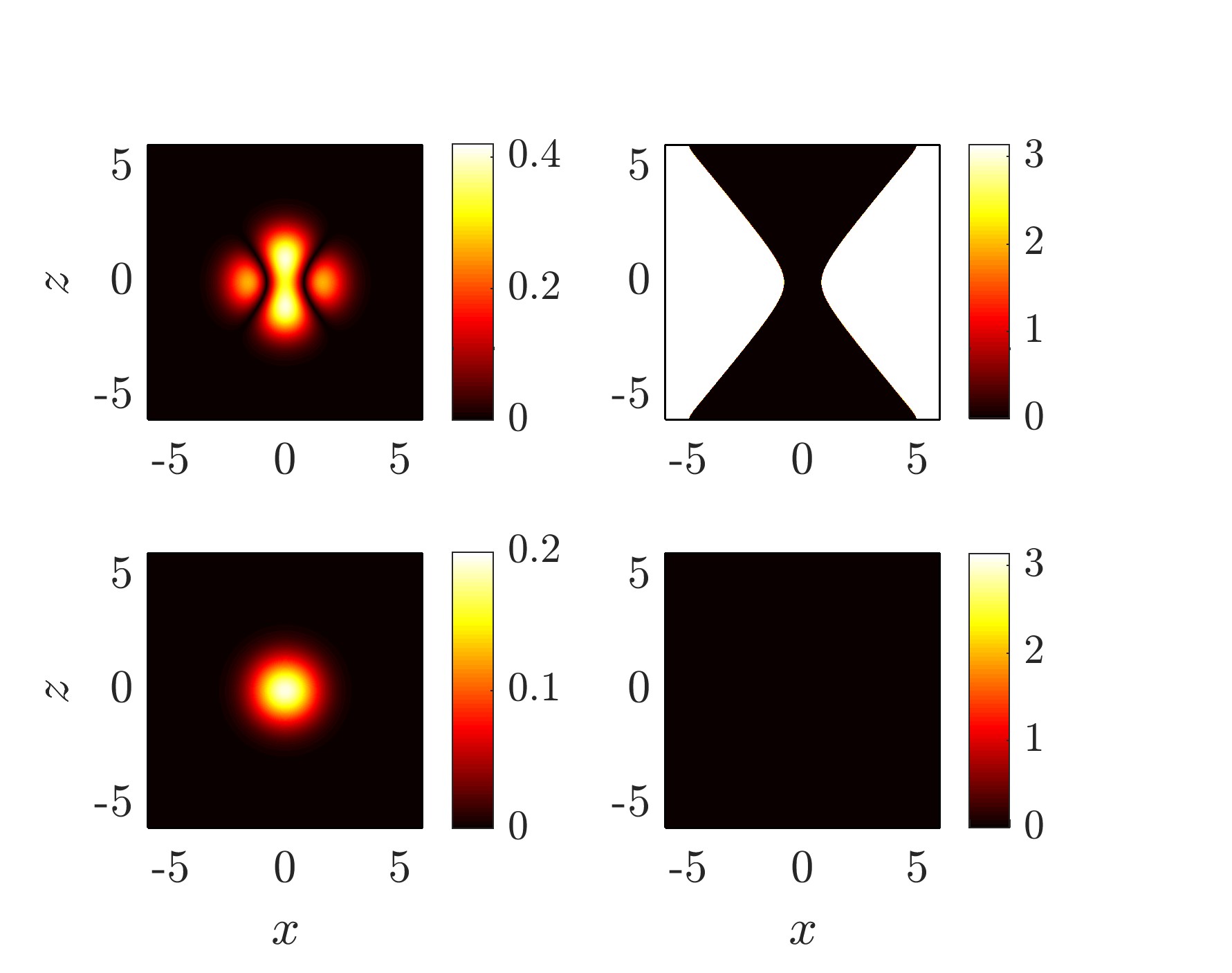}
\includegraphics[width=\columnwidth]{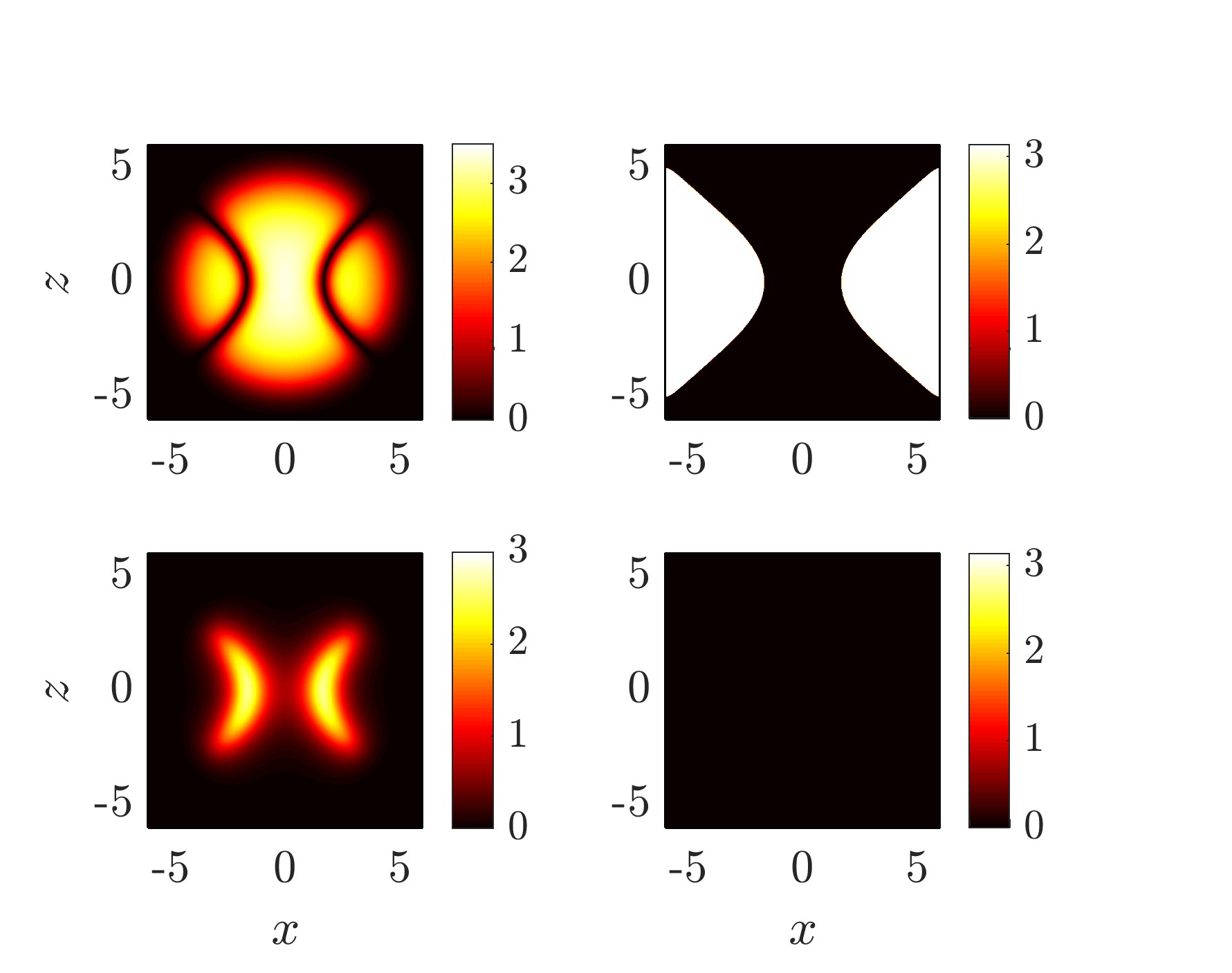}
\caption{Stationary states of the RDB soliton near the linear limit $\mu_1=3.55, \mu_2=1.55$ (top panels) and in the Thomas-Fermi limit $\mu_1=12, \mu_2=10$ (lower panels). Note the bright component has the same phase throughout the condensate, and is trapped by the RDS at high densities.
}
\label{states2}
\end{figure}

\begin{figure}%{r}{0.5\textwidth}
\includegraphics[width=\columnwidth]{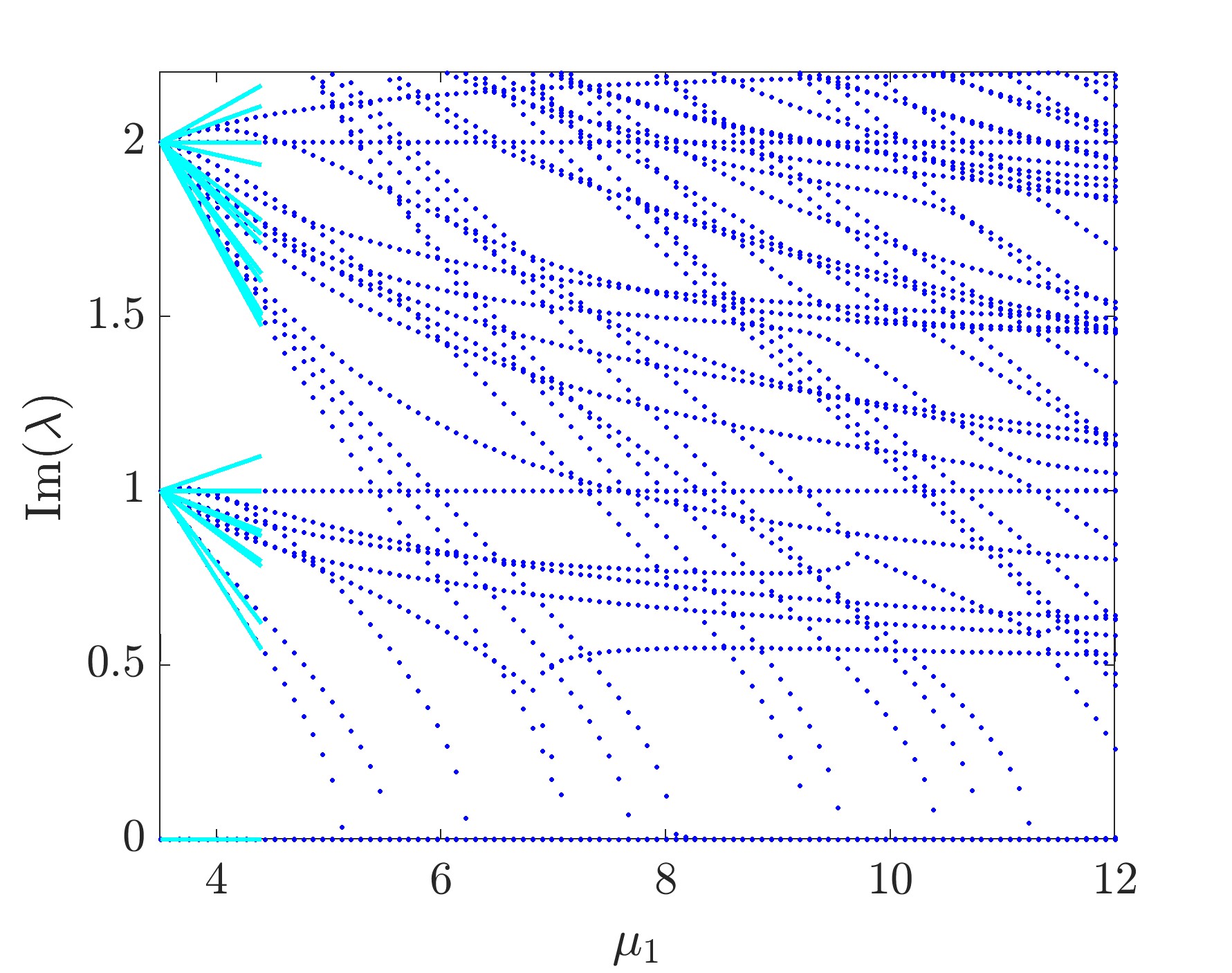}
\includegraphics[width=\columnwidth]{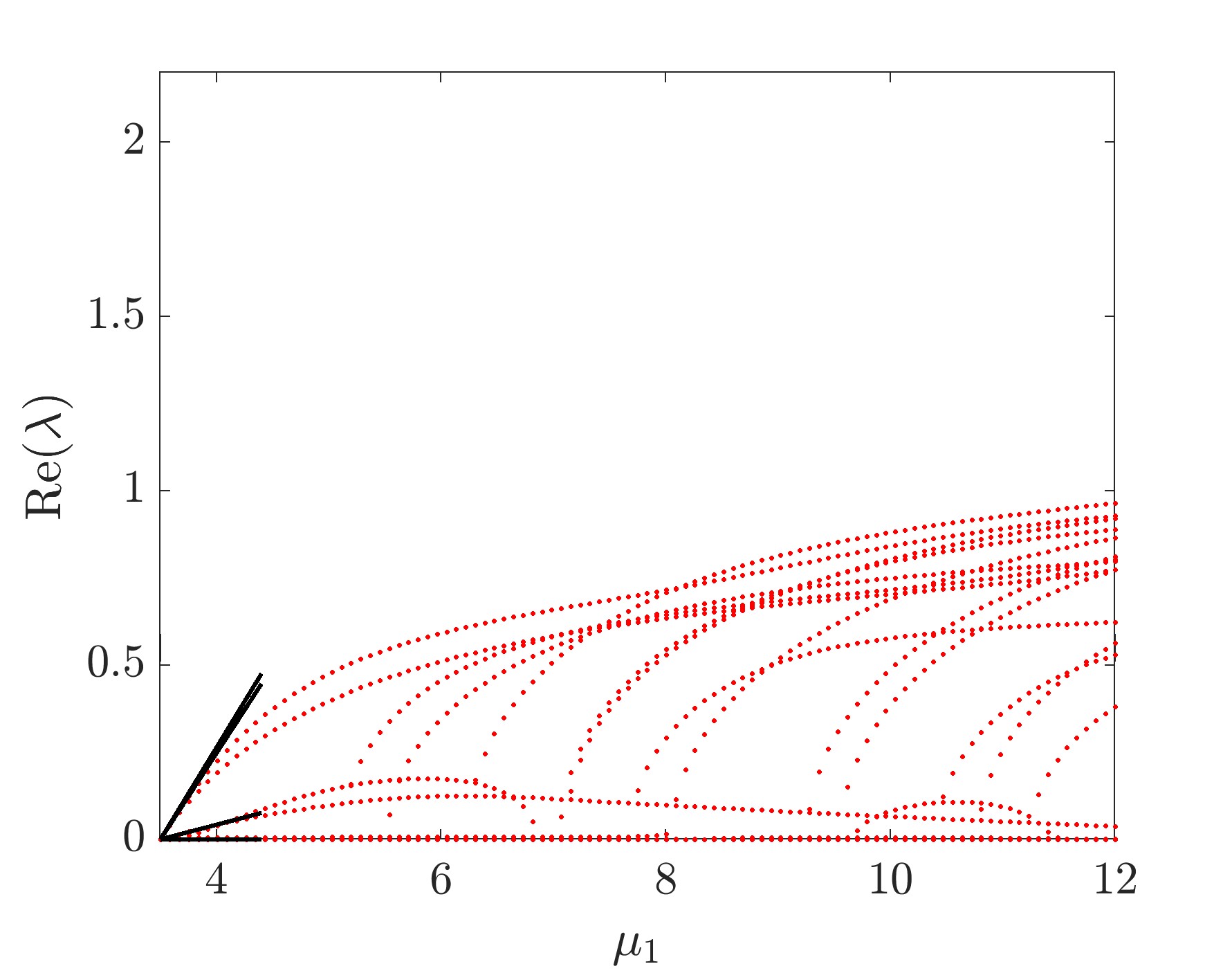}
\caption{The spectrum of the RDB soliton along with the NLLDPM analysis from the linear limit to a typical Thomas-Fermi limit $\mu_1=12, \mu_2=10$ with a linear ``trajectory'' in the ($\mu_1,\mu_2$) parameter space, hence only the variable $\mu_1$ is shown. The RDS is much more stable in the presence
  of the bright component; compare with Fig.~\ref{spectrum}. However, the RDS still becomes unstable right from the linear limit.
}
\label{spectrum2}
\end{figure}

Two stationary states, one near the linear limit and one in the Thomas-Fermi limit, are shown in Fig.~\ref{states2} and the spectrum for the aforementioned parameters along with the modified semi-analytical spectrum is shown in Fig.~\ref{spectrum2}.
%It is natural that the bright component is strongly coupled to the RDS in the Thomas-Fermi limit.
The RDB soliton is much less unstable than the single-component RDS; compare with Fig.~\ref{spectrum}. This is both in terms of the number of unstable modes and the maximum growth rate. The RDB is still unstable right from the linear limit, but it should be much easier to observe long-lived RDB solitons in
two-component systems, given their considerably weaker instabilities.

%One should, however, not conclude such stabilization by a bright component is impossible. It is perhaps hard to stabilize the RDS when $\mu_1$ and $\mu_2$ are very comparable. There is still a possibility to fix $\mu_2$ and start decreasing $\mu_1$, i.e., using a high-density bright component to stabilize a low-density RDS \cite{Panos:RDB}. It seems likely that such a procedure is rather similar to the idea of applying an external stabilization potential. Therefore, we explore this easier mechanism for the single-component RDS.

\begin{figure}%{r}{0.5\textwidth}
\includegraphics[width=\columnwidth]{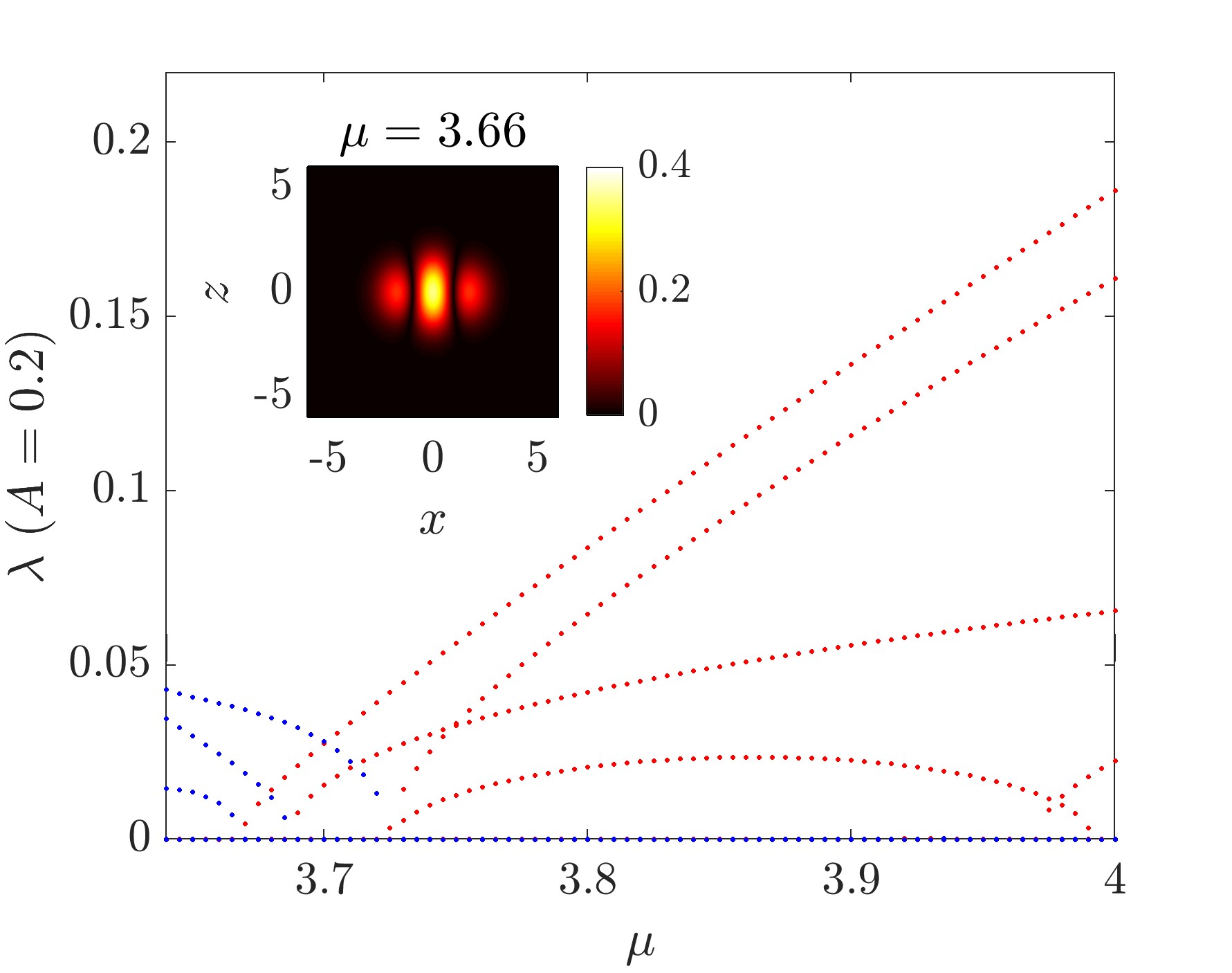}
\includegraphics[width=\columnwidth]{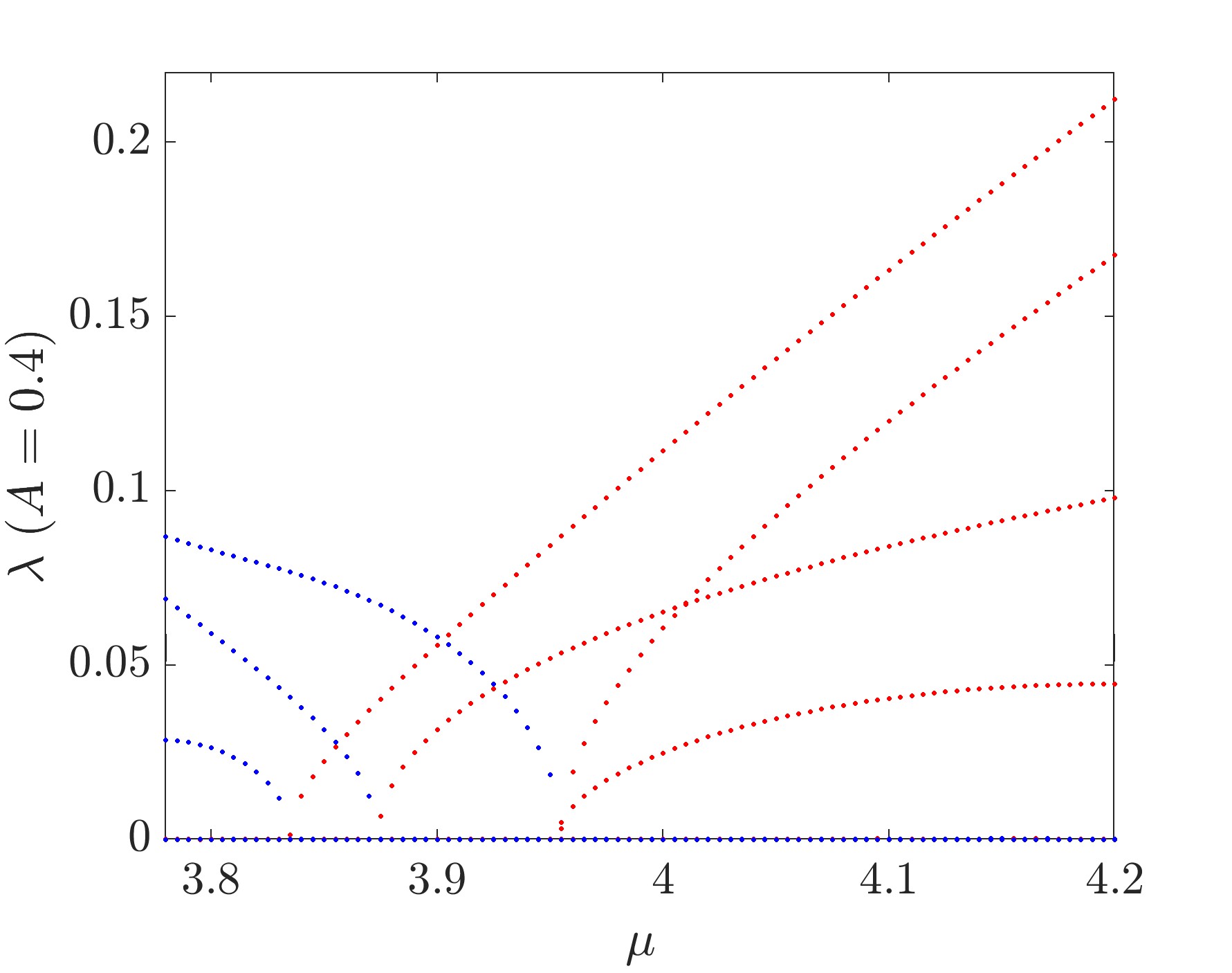}
\caption{The spectrum of the RDS with an additional repulsive ring potential (Eq.~(\ref{Vp})) around the ring. The top panel has a potential strength $A=0.2$, while the bottom panel has a stronger strength $A=0.4$ with a wider range of stability, note the different $\mu$ scales. The top inset panel shows a stable RDS density profile. The RDS is fully stabilized near the linear limit, and the RDS also gets more straight with the applied vertical potential. The stability is verified dynamically.
}
\label{spectrum3}
\end{figure}

Our second attempt towards stabilization of a RDS
is to use an external potential acting on the RDS, which can be realized by the blue-detuned laser \cite{BDlaser}. This has been used in two-dimensional systems to stabilize the dark soliton stripe and also more importantly the dark soliton ring \cite{Ma:DS,Wang:DSR}.
Now we study a potential of the following form
\begin{eqnarray}
V=\frac{1}{2} \omega_{\rho}^2 \rho^2+\frac{1}{2} \omega_z^2 z^2+A\exp[-(\rho-\rho_0)^2/(2\sigma^2)],
\label{Vp}
\end{eqnarray}
where for simplicity we set $\rho_0=1$ as we are mostly interested
in the vicinity of the linear limit here. We take typical values $A=0.2$ and $\sigma=1$ unless specified otherwise.

The repulsive potential typically delays the existence of the RDS in terms of the chemical potential. For our particular choice of the potential, the RDS starts to exist from about $\mu=3.64$, compared with $\mu=3.5$ in the harmonic trap. The spectrum is shown in Fig.~\ref{spectrum3}. Indeed, we find
the RDS is fully stabilized near the linear limit up to about $\mu=3.665$. The
RDS is also a lot more straight as a result of the added vertical ring potential.

We have performed direct numerical simulations of the evolution
of the RDS for a parameter value of chemical potential $\mu=3.66$
such that the RDS should be stabilized. 
Indeed, we have verified its stability up to $t=1000$. In addition, we have also implemented a stationary state with random noise of both signs of strength
about $5\%$ of the maximum field absolute value, and again we have
confirmed its stability up to $t=1000$. This serves to point out that
the stabilization of the RDS is feasible in the presence
of an external barrier potential even if for a narrow parametric range.
One can envision modifying the potential parameters to expand
upon this range, yet we believe that the above computations illustrate
the proof-of-principle of the relevant idea. We have therefore explored an additional example using a stronger potential $A=0.4$. The results are qualitatively similar and indeed give a larger stability range as shown in the bottom panel of Fig.~\ref{spectrum3}. Here, the existence of the RDS is naturally delayed further to about $\mu=3.78$. The instabilities are further suppressed and the RDS in this case is stable up to about $\mu=3.83$.
It should be noted here that the stabilization of the RDS structures by a cylindrical external potential of the form of Eq.~(\ref{Vp}) is not a priori guaranteed as the RDS structure bends. Yet, our results clearly illustrate that it is possible.

%since the full potential stemming from
%the combination of Eq.~(\ref{potential}) and~(\ref{Vp}) does not have a
%definite symmetry. 

Similarly, one can use the same method to stabilize the RDB soliton. For $A=0.2$, the state is found to exist along the linear parametric line from $(3.65, 1.7)$ to $(4,2.05)$ and the state is stable again from the linear limit up to $\mu_1=3.665$. For $A=0.4$, the state exists from $(3.79, 1.89)$ to $(4.2,2.3)$ and the state is stable up to $\mu_1=3.835$, i.e., bearing a slightly larger
  stability range. We have performed direct numerical simulations
  and confirmed the stability of the structure up to $t=1000$ (results not shown here).
  We have also conducted a typical SO(2) rotation which induces breathing dynamics in the Manakov model; see, e.g., Ref.~\cite{Wang:SO2},
where this is illustrated both for one-dimensional dark-dark soliton states,
but also in two-dimensional, so-called vortex-vortex states.
  A relevant example is shown for $\mu_1=3.83, \mu_2=1.93$ in
  Ref.~\cite{RDSmv4} featuring a
  movie of the stable breathing pattern between two ring dark solitons in the
  two components (a RDS-RDS state).

\section{Conclusions and Future Challenges}
\label{conclusion}
In this work, we have presented a systematic study of
the extension of the cylindrical ring dark soliton structure
in three-dimensional Bose-Einstein condensates. We have studied its existence, stability and dynamics. The RDS has been identified
as possessing numerous unstable modes, and typical destabilization dynamics generates more robust vortical structures, such as the two or four parallel vortex rings with alternating vorticity. Despite the fact that the
RDS is highly unstable, we have demonstrated that it can be stabilized with a suitable ring-shaped repulsive potential to trap the RDS. In the process
of exploring the unstable spectral modes of the structure,
we have also provided a hybrid (between numerical and analytical) procedure
for extrapolating the modes of the linearization based on a few computations
in the vicinity of the linear limit and utilizing the corresponding
theoretical setup. This methodology is especially well suited, as discussed
in the text, for structures that may suffer reduced symmetry
such as the RDS or the vortex ring in a spherical (or more generally
ellipsoidal) trap.

The methods developed or summarized in this work can be readily applied to a wide range of solitary waves in three-dimensional condensates in both one- and two-component systems. Some interesting examples include the vortex-line-bright and the vortex-ring-bright structures \cite{Wang:SO2}. Other dark soliton surfaces and vortical filaments are also interesting, and their systematic studies are useful towards formulating effective filament-level dynamical theories for pattern formation within Bose-Einstein condensates~\cite{Wang:VR,Horng:VR}.

\begin{acknowledgments}
We thank B.A.Malomed and R. Carretero-Gonz{\'a}lez for useful feedback on the manuscript. 
W.W. and E.B. acknowledge support from the Swedish Research Council Grant No.~642-2013-7837 and Goran Gustafsson Foundation for Research in
Natural Sciences and Medicine. This material is based upon work supported by the U.S.
National Science Foundation under Grant No. PHY-1602994
(P.G.K.).
The computations were performed on resources
provided by the Swedish National Infrastructure for Computing (SNIC)
at the National Supercomputer Centre (NSC) and the High Performance Computing Center North (HPC2N).

\end{acknowledgments}

\bibliography{Refs}

\end{document}